\documentclass[%
 pra,
 superscriptaddress,
 twocolumn,
 longbibliography,   % ★ 추가
 amsmath,amssymb,
 aps,
]{revtex4-2}
\usepackage{graphicx}% Include figure files
\usepackage{svg} % Support for .svg images
\usepackage{dcolumn}% Align table columns on decimal point
\usepackage{bm}
\usepackage[colorlinks=true,allcolors=blue]{hyperref}% bold math
\newcommand{\ket}[1]{\left|#1\right\rangle}

\begin{document}

\preprint{APS/123-QED}

\title{A cavity-mediated reconfigurable coupling scheme for superconducting qubits}% Force line breaks with \\
% A tunable (Reconfigurable / Programmable + long-distance) all-to-all / coupling scheme 
% A tunable long-range coupling scheme for superconducting quantum processors
% A tunable, photon(resonator, cavity)-mediated coupling scheme for superconducting quantum processors

\author{Shinyoung Hwang}
\altaffiliation{Current address: Pritzker School of Molecular Engineering, University of Chicago, Chicago, IL 60637, USA}
\affiliation{Institute of Applied Physics, Seoul National University, Seoul 08826, Korea}
  \affiliation{NextQuantum Innovation Research Center, Seoul National University, Seoul 08826, Korea}
  
\author{Sangyeon Lee}
  \affiliation{Institute of Applied Physics, Seoul National University, Seoul 08826, Korea}
  \affiliation{NextQuantum Innovation Research Center, Seoul National University, Seoul 08826, Korea}
\affiliation{Department of Physics and Astronomy, Seoul National University, Seoul 08826, Korea}
\author{Eunjong Kim}
 \email{eunjongkim@snu.ac.kr}
  \affiliation{Institute of Applied Physics, Seoul National University, Seoul 08826, Korea}
  \affiliation{NextQuantum Innovation Research Center, Seoul National University, Seoul 08826, Korea}
   \affiliation{Department of Physics and Astronomy, Seoul National University, Seoul 08826, Korea}

\date{\today}% It is always \today, today,

\begin{abstract}
% 시작 부분은 soft하게 시작. 
Superconducting qubits have achieved remarkable progress in gate fidelity and coherence, yet their typical nearest-neighbor connectivity presents constraints for implementing complex quantum circuits. Here, we introduce a cavity-mediated coupling architecture in which a shared cavity mode, accessed through tunable qubit–cavity couplers, enables dynamically reconfigurable interactions between non-adjacent qubits. By selectively activating the couplers, we demonstrate that high-fidelity iSWAP and CZ gates can be performed within 50\:ns with simulated coherent error below $10^{-4}$, while residual $ZZ$ interaction during idling remains below a few kilohertz. Extending to a four-qubit system, we also simulate gates between every qubit pair by selectively enabling the couplers with low qubit crosstalk. This approach provides a practical route toward enhanced interaction flexibility in superconducting quantum processors and may serve as a useful building block for devices that benefit from selective non-local coupling.   % incoherent error가 있을때의 gate fidelity estimate. %+ simultaneous 한것?  %+ 이 연구의 potential, broader impact
\end{abstract}

%\keywords{proposeed keywords}%Use showkeys class option if keyword

%\tableofcontents
\maketitle
\section{Introduction}
% Paragraph 1
Quantum computers promise computational advantages over classical methods across diverse domains, including physics, chemistry, and combinatorial optimization \cite{1996lloyd, 2005aspuru, 1999shor, Grover1996, 2022opti}. Realizing these advantages depends critically on the connectivity of the underlying qubit architecture, which determines the algorithmic efficiency and the range of interactions that can be implemented. Superconducting quantum processors, however, have predominantly adopted local connectivity layouts, most notably square lattices \cite{2022surfacecodeQEC, 2023scaledsurfacecode} and heavy-hex geometries \cite{2023scalableerrormitigation,2025ibmreview}. While these nearest-neighbor architectures have enabled major advances in quantum error correction \cite{2025belowthreshold} and quantum simulation \cite{2023utility, 2025fermiHubbard72, 2025Z2gaugestrings, 2025quantumergodicityedge}, operations between distant qubits require swap networks \cite{2019swapnet} that increase circuit depth and accumulated error. From a theoretical perspective, non-local interactions enable quantum error-correcting codes with higher encoding rates and reduced overhead \cite{2022thinplanar, 2024ldpc, 2024constantoverheadFT}, and allow the native realization of long-range-interacting many-body systems for quantum simulation \cite{2013frustration, 2016competingrange, 2021emergentgeometry, 2023metamaterial, 2021MBLwithLR, 2020scramble}.

% Paragraph 2
Extending qubit connectivity beyond nearest neighbors in superconducting circuits remains challenging due to the intrinsically local nature of capacitive and inductive couplings. One widely adopted approach therefore uses transmission-line resonators as quantum buses \cite{2004bus} to mediate long-range interactions, enabling multi-qubit entanglement and resonator-based connectivity networks \cite{2017tenqubitbus, 201920qubitbus, demoldpc}. However, scaling such bus-mediated architectures introduces fundamental limitations: longer resonators or resonators coupled to many qubits lead to spectral crowding of modes, complicating frequency allocation and control, while shared mediating modes produce partially always-on interactions that lack independent tunability, resulting in crosstalk and leakage. Alternative approaches have explored multi-step qubit–resonator gate sequences \cite{2025eff} or large-capacitance router nodes \cite{2024router}, trading improved gate performance for increased operational complexity or limited coupling range.

% Paragraph 3
In this work, we propose a cavity-mediated, dynamically reconfigurable coupling architecture for superconducting qubits that enables interactions between arbitrary qubit pairs through a shared bus resonator controlled by frequency-tunable couplers. In the idle configuration, qubits are effectively decoupled from the resonator, while selected qubit-qubit interactions can be activated on demand without direct capacitive coupling of qubits. This relaxes geometric constraints on qubit placement and distinguishes our approach from conventional tunable coupler designs \cite{2018tunableCoupler, 2021tunable}. Numerical simulations show that the architecture achieves strong effective coupling with excellent idling isolation, enabling iSWAP and CZ gates within 50 ns with coherent errors below $10^{-4}$ and residual $ZZ$ interactions suppressed to the kilohertz level. We further demonstrate selective non-local two-qubit gates in multi-qubit settings with minimal crosstalk, highlighting the scalability of the approach.

\begin{figure*}[t]  % figure* → 두 컬럼 전체 폭 사용
    \centering
    \includegraphics[width=0.95\textwidth]{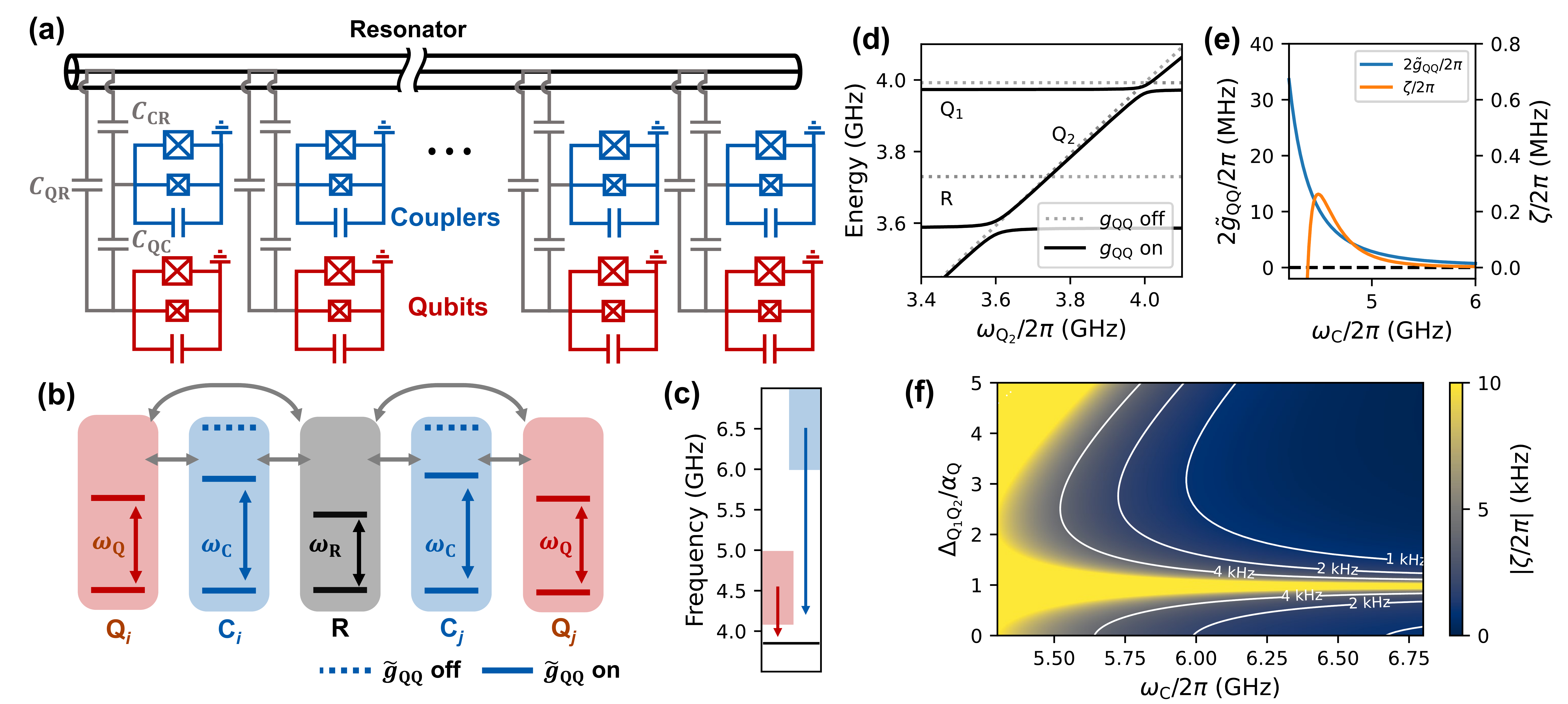}
    \caption{\label{fig:fig1} 
    Architecture for reconfigurable connectivity of superconducting qubits. (a) Circuit schematic showing qubits (red), tunable couplers (blue), and a shared bus resonator (black), coupled via $C_{\mathrm{QC}}$ (qubit-coupler), $C_{\mathrm{QR}}$ (qubit-resonator), and $C_{\mathrm{CR}}$ (coupler-resonator). (b) Single-excitation level diagram for a selected qubit pair ($\mathrm{Q}_i$, $\mathrm{Q}_j$), with couplers active (solid) or idle (dashed).  (c) Operating frequency ranges of the qubits, couplers, and the fundamental resonator mode (black). In the idle configuration, modes are detuned to suppress interactions and brought closer to activate coupling. (d) Single-excitation spectrum versus $\omega_{\mathrm{Q}_2}$, where an avoided crossing appears only when the couplers are enabled. (e) Effective exchange coupling $\tilde{g}_{\mathrm{QQ}}$ and residual $ZZ$ interaction $\zeta$ versus coupler frequency $\omega_{\mathrm{C}}$ with the qubits on resonance. (f) Residual $ZZ$ interaction versus coupler frequency $\omega_\mathrm{C}$ and qubit detuning $\Delta_{\mathrm{Q_1Q_2}}$ ($\mathrm{Q}_2$ fixed at $4$\:GHz), demonstrating robust suppression over a broad parameter range.
    }
\end{figure*}

\section{Architecture and Operating Principle}

% Paragraph 5
Our proposed architecture, illustrated in Fig.~\ref{fig:fig1}(a), realizes effective qubit-qubit interactions through a shared cavity mode implemented as a $\lambda/2$ coplanar waveguide resonator (black, labeled R with frequency $\omega_\mathrm{R}$).  
The computational units are tunable-frequency transmon qubits (red, labeled $\mathrm{Q}_i$ with frequencies $\omega_{\mathrm{Q}_i}$) that are capacitively coupled to the resonator.  
Each qubit is connected to the cavity via an intermediate frequency-tunable, transmon-like coupler (blue, labeled $\mathrm{C}_i$ with frequency $\omega_{\mathrm{C}_i}$), which provides a controllable interaction pathway between the qubit and the resonator \cite{2018tunableCoupler, 2021tunable}.  
The coupling capacitances between qubit and coupler ($C_{\mathrm{QC}}$), and between coupler and resonator ($C_{\mathrm{CR}}$), are engineered to yield a large dynamic range in the effective qubit-resonator interaction, while a small direct qubit-resonator capacitance ($C_{\mathrm{QR}}$) remains.

% Paragraph 6
Because the qubit-resonator interaction is mediated by tunable couplers, the effective coupling strength is controlled by the coupler frequency [Fig.~\ref{fig:fig1}(b)]. By biasing the couplers far detuned, the system supports an idle configuration in which all qubits are effectively decoupled from the cavity, and consequently, from one another. This isolation does not rely on direct qubit-qubit capacitance or interference between multiple pathways and therefore imposes no intrinsic constraint on qubit spacing beyond the resonator length. The architecture thus enables selective interactions between arbitrary qubit pairs while maintaining strong spectator isolation. The corresponding bare frequencies are summarized in Fig.~\ref{fig:fig1}(c): in the idle configuration, qubits are parked in the 4–5\:GHz range and couplers at higher frequencies (6–7\:GHz). Lowering the relevant couplers to $\sim4.3$\:GHz activates the interaction, while the target qubits are tuned near 4\:GHz, slightly above the resonator frequency.

% Paragraph 7
When an interaction between a selected pair of qubits $\mathrm{Q}_i$ and $\mathrm{Q}_j$ is enabled, the cavity mediates an effective qubit-qubit coupling $\tilde{g}_{\mathrm{QQ}}$ through the corresponding couplers, while all other qubits remain spectrally isolated.  
This mediated interaction supports fast, non-adiabatic two-qubit gates without requiring direct qubit-qubit coupling.  
An iSWAP gate is realized by bringing the two target qubits into resonance and allowing coherent population exchange between the $\ket{10}$ and $\ket{01}$ states.  
A CZ gate is obtained by also using population oscillations, but now between the $\ket{11}$ state and second-excited state of one qubit, $\ket{20}$ or $\ket{02}$. By completing a single period of the oscillation, the $\ket{11}$ acquires a controlled phase desired for the CZ gate.
In both cases, the interaction is mediated entirely by the cavity mode and can be selectively applied to any qubit pair supported by the architecture.

\section{Effective Hamiltonian and Qubit Coupling}

% Paragraph 8
We model the proposed architecture by formulating its circuit Hamiltonian and extracting the resulting qubit-qubit coupling through numerical diagonalization. Each qubit and coupler is treated as a weakly anharmonic Duffing oscillator, while the resonator is modeled as a harmonic oscillator. The resonator is treated as a single harmonic mode, with higher-order spatial harmonics neglected.

% Paragraph 9
The total Hamiltonian is written as $H_{\mathrm{tot}} = H_{\mathrm{self}} + H_{\mathrm{int}}$, where
\begin{equation}
H_{\mathrm{self}} =
\sum_{m,i}
\left(
\omega_{m_i} a_{m_i}^\dagger a_{m_i}
+ \frac{\alpha_{m_i}}{2}
a_{m_i}^\dagger a_{m_i}^\dagger a_{m_i} a_{m_i}
\right),
\label{eq:self-H}
\end{equation}
and
\begin{eqnarray}
H_{\mathrm{int}}
&=&
-\sum_i \sum_{\substack{m,n \\ m \neq n}}
g_{m_i n_i}
\left(a_{m_i}^\dagger - a_{m_i}\right)
\left(a_{n_i}^\dagger - a_{n_i}\right)
\nonumber \\
&&
-\sum_{\substack{i,j \\ i>j}}
g_{\mathrm{C}_i \mathrm{C}_j}
\left(a_{\mathrm{C}_i}^\dagger - a_{\mathrm{C}_i}\right)
\left(a_{\mathrm{C}_j}^\dagger - a_{\mathrm{C}_j}\right),
\label{eq:int-H}
\end{eqnarray}
with circuit quantization detailed in App.~\ref{sec:quantization}. Here $m,n \in \{\mathrm{Q},\mathrm{C},\mathrm{R}\}$ label qubit, coupler, and resonator modes, and $i,j$ index qubit-coupler units. The operators $a_{m_i}$, $\omega_{m_i}$, and $\alpha_{m_i}$ denote the annihilation operator, frequency, and anharmonicity of each mode, with $\alpha_{\mathrm{R}}=0$ for the resonator. The first term in $H_{\mathrm{int}}$ describes interactions within each unit, while the second captures weak coupler-coupler couplings mediated by the capacitance network. For simplicity, uniform capacitances are assumed:
$C_{\mathrm{Q}_i\mathrm{C}_i}=C_{\mathrm{QC}}$,
$C_{\mathrm{C}_i\mathrm{R}}=C_{\mathrm{CR}}$, and
$C_{\mathrm{Q}_i\mathrm{R}}=C_{\mathrm{QR}}$.

To obtain an effective description, we eliminate the coupler degrees of freedom using a Schrieffer–Wolff transformation, which renormalizes the frequencies and generates effective qubit–resonator couplings. 
%Because the couplers are mutually far detuned and weakly coupled ($|\Delta_{\mathrm{C}_i\mathrm{C}_j}|\gg g_{\mathrm{CC}}$), coupler–coupler interactions are neglected to leading order.
After the rotating-wave approximation, the Hamiltonian reduces to
$\tilde{H}_{\mathrm{eff}}=\tilde{H}_{\mathrm{self}}+\tilde{H}_{\mathrm{int}}$, with
\begin{align}
\tilde{H}_{\mathrm{self}} &=
\sum_i
\left(
\tilde{\omega}_{\mathrm{Q}_i} a_{\mathrm{Q}_i}^\dagger a_{\mathrm{Q}_i}
+
\frac{\alpha_{\mathrm{Q}_i}}{2} a_{\mathrm{Q}_i}^{\dagger 2} a_{\mathrm{Q}_i}^{2}
\right)
+
\tilde{\omega}_{\mathrm{R}} a_{\mathrm{R}}^\dagger a_{\mathrm{R}}, \\
\tilde{H}_{\mathrm{int}} &=
\sum_i
\tilde{g}_{\mathrm{Q}_i\mathrm{R}}
\left(
a_{\mathrm{Q}_i}^\dagger a_{\mathrm{R}}
+
a_{\mathrm{Q}_i} a_{\mathrm{R}}^\dagger
\right),
\end{align}
describing qubits coupled to a common resonator with tunable effective couplings. Here, the renormalized parameters are
\[
\tilde{\omega}_{\mathrm{Q}_i}
=\omega_{\mathrm{Q}_i}+\frac{g_{\mathrm{Q}_i\mathrm{C}_i}^2}{\Delta_{\mathrm{Q}_i\mathrm{C}_i}},
\quad
\tilde{\omega}_{\mathrm{R}}
=\omega_{\mathrm{R}}+\sum_i\frac{g_{\mathrm{C}_i\mathrm{R}}^2}{\Delta_{\mathrm{R}\mathrm{C}_i}},
\]
\[
\tilde{g}_{\mathrm{Q}_i\mathrm{R}}
=\frac{g_{\mathrm{Q}_i\mathrm{C}_i} g_{\mathrm{C}_i\mathrm{R}}}{\Delta_i}
+ g_{\mathrm{Q}_i\mathrm{R}},
\]
where $\Delta_{m_i n_j}=\omega_{m_i}-\omega_{n_j}$ and
$\Delta_i^{-1}=\tfrac12(\Delta_{\mathrm{Q}_i\mathrm{C}_i}^{-1}+\Delta_{\mathrm{R}\mathrm{C}_i}^{-1})$.
Tuning the couplers so that only selected $\tilde g_{\mathrm{Q}_i\mathrm{R}}$ are appreciable activates interactions between chosen qubit pairs while leaving the others effectively idle (App.~\ref{sec:frequency-config}).

% Paragraph 12
We first characterize the mediated interaction using numerical diagonalization of the full Hamiltonian in Eqs.~\eqref{eq:self-H}--\eqref{eq:int-H}, considering the minimal case of two qubits ($\mathrm{Q}_1$ and $\mathrm{Q}_2$) coupled to the cavity ($\mathrm{R}$). 
Figure~\ref{fig:fig1}(d) shows the single-excitation spectrum as the $\mathrm{Q}_2$ frequency is swept with the couplers either activated or detuned to the idle configuration. The qubits are centered at $\omega_{\mathrm{Q}_{1,2}}/2\pi = 4.0$\:GHz and coupled to a resonator at $\omega_{\mathrm{R}}/2\pi = 3.85$\:GHz. When the couplers are activated, an avoided crossing appears at resonance, and the resulting level splitting defines the effective exchange coupling $\tilde{g}_{\mathrm{QQ}}$; in the idle configuration, the spectrum remains crossing, indicating negligible coupling.

\begin{figure}[b!]  % figure* 
    \centering
    \includegraphics[width=0.95\linewidth]{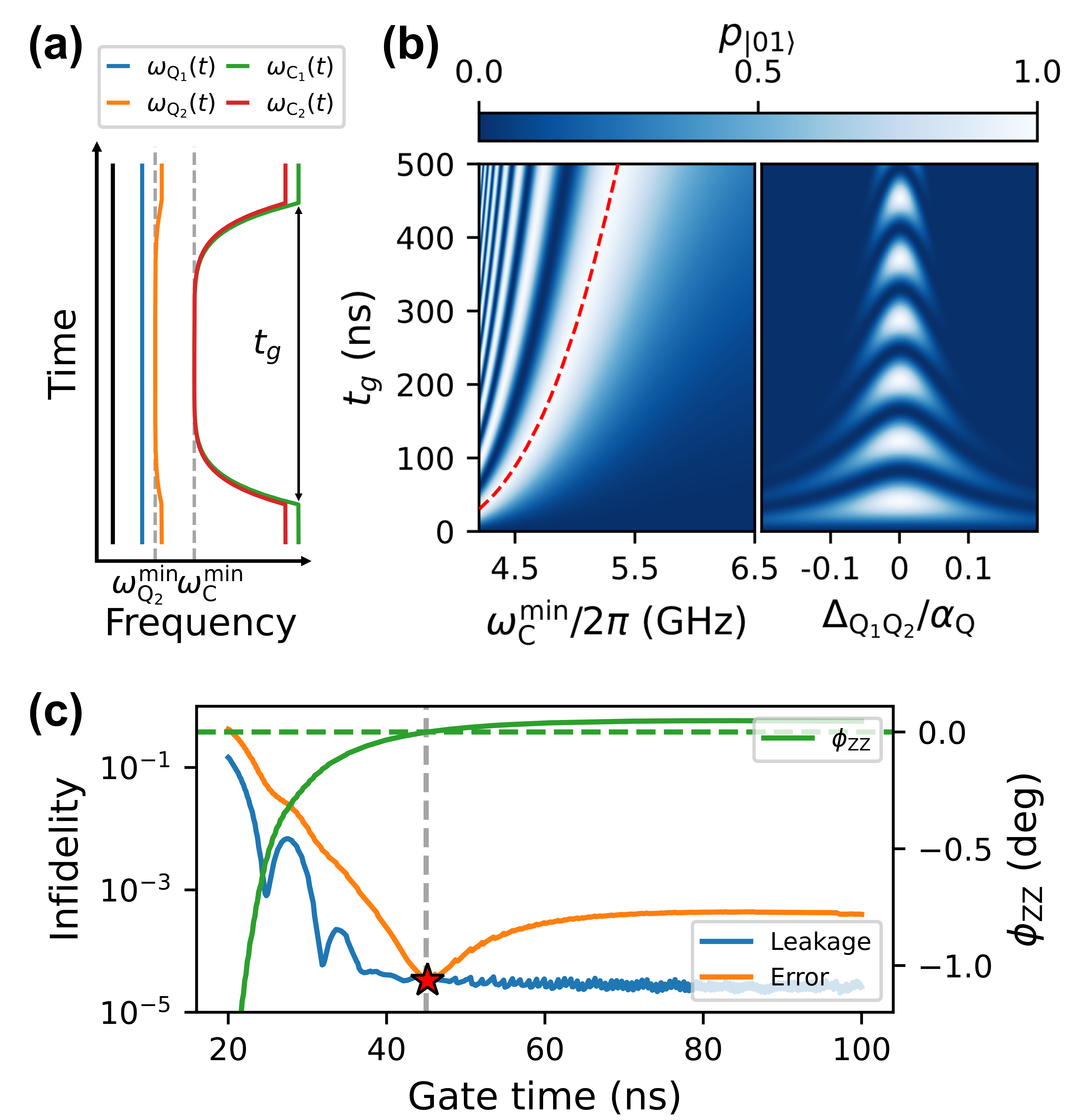}
    \caption{\label{fig:fig2}
    Time evolution for activating the effective exchange coupling and simulated iSWAP gate errors. 
    (a) Frequency trajectories during a gate of duration $t_g$: the two couplers are tuned to $\omega_{\mathrm C}^{\min}$ and $\mathrm{Q}_2$ to $\omega_{\mathrm Q_2}^{\min}$, while $\mathrm{Q}_1$ and the resonator (black) remain fixed. 
    (b) Vacuum Rabi oscillations between $\ket{10}$ and $\ket{01}$ as a function of coupler minimum frequency $\omega_{\mathrm C}^{\min}$ at fixed $\Delta_{\mathrm{Q_1Q_2}}=0$ (left) and as a function of qubit detuning $\Delta_{\mathrm{Q_1Q_2}}$ at fixed $\omega_{\mathrm C}^{\min}$ (right); the red dashed line indicates the operating trajectory used to implement the iSWAP gate. (c) Simulated iSWAP errors: leakage (blue), total infidelity (orange), and residual controlled phase $\phi_{ZZ}$ (green).
}
\end{figure}

% Paragraph 13–
Figure~\ref{fig:fig1}(e) summarizes the dependence of $\tilde{g}_{\mathrm{QQ}}$ and the residual $ZZ$ interaction on the coupler frequency. For $(C_{\mathrm{QC}}, C_{\mathrm{QR}}, C_{\mathrm{CR}})=(3.7,1.0,37.4)$\:fF and anharmonicities $(\alpha_{\mathrm{Q}}/2\pi, \alpha_{\mathrm{C}}/2\pi)=-(220,268)$\:MHz, we obtain $2\tilde{g}_{\mathrm{QQ}}/2\pi \approx 25$\:MHz near $\omega_{\mathrm{C}}/2\pi = 4.3$\:GHz, corresponding to an iSWAP time of $\sim 45$\:ns. As the coupler frequency increases, the exchange interaction decreases smoothly and vanishes in the far-detuned regime. 
The residual $ZZ$ interaction remains small ($<5$\:kHz) across the entire operating range  [Fig.~\ref{fig:fig1}(f)], indicating robust suppression of parasitic coupling.

% Paragraph 15
\section{Gate operation}\label{sec:gate-operations}

We now consider time-dependent control of the qubit and coupler frequencies, $\omega_{\mathrm{Q}_i}(t)$ and $\omega_{\mathrm{C}_i}(t)$, to implement effective two-qubit operations $U$. Gate performance is evaluated by simulating the unitary evolution and optimizing control parameters to approximate the targeted ideal gates $U_{\mathrm{id}}$, from which average gate fidelities are extracted \cite{dtcfidel}.

Polynomial-shaped, Slepian-like trajectories are used for frequency control to achieve low leakage to coupler states \cite{2021tunable},

\[
\omega_{m_i}(t)
=
\omega_{m_i}(0)
-\Delta_{m_i}\left[1-(2t/t_g-1)^{2n}\right],
\]
where 
$m\in\{\mathrm{Q},\mathrm{C}\}$, $t_g$ is the duration of the gate, $\Delta_{m_i}$ the maximum excursion, and $n$ the pulse order. As a representative example, we simulate vacuum Rabi oscillations between $\ket{10}$ and $\ket{01}$ using the trajectories shown in Fig.~\ref{fig:fig2}(a). Bringing the two qubits into resonance while tuning the couplers to a common minimum frequency $\omega_\mathrm{C}^{\min}$ activates the exchange interaction, producing coherent population transfer [left panel of Fig.~\ref{fig:fig2}(b)]. Sweeping the qubit detuning yields the characteristic chevron pattern in the right panel of Fig.~\ref{fig:fig2}(b).

\subsection{iSWAP gate}
An iSWAP gate is realized by completing a full population exchange between the two qubits, corresponding to
\[
\int_0^{t_g} \tilde{g}_{\mathrm{QQ}}(t)\, dt = \pi .
\]
Gate fidelities are evaluated following the procedure of App.~\ref{sec:gate-fidelity}. As shown in Fig.~\ref{fig:fig2}(c), leakage errors below $10^{-4}$ are achieved for $t_g \gtrsim 35$\:ns, indicating that strong cavity-mediated coupling enables fast operations with minimal population loss.

In addition to leakage, coherent errors arise from the residual $ZZ$ interaction $\zeta$, which accumulates a residual controlled phase $\phi_{ZZ}=\int^{t_g}_0 \zeta(t) dt$ during the gate. Because $\zeta$ changes sign as a function of the coupler frequency [Fig.~\ref{fig:fig1}(e)], positive and negative contributions can be balanced to cancel the net phase \cite{2021tunable}. At the optimal operating point [red star in Fig.~\ref{fig:fig2}(d)], the accumulated $\phi_{ZZ}$ vanishes, yielding a $ZZ$-free iSWAP gate whose fidelity is limited primarily by leakage.

\begin{figure}[t]
    \centering
    \includegraphics[width=0.95\linewidth]{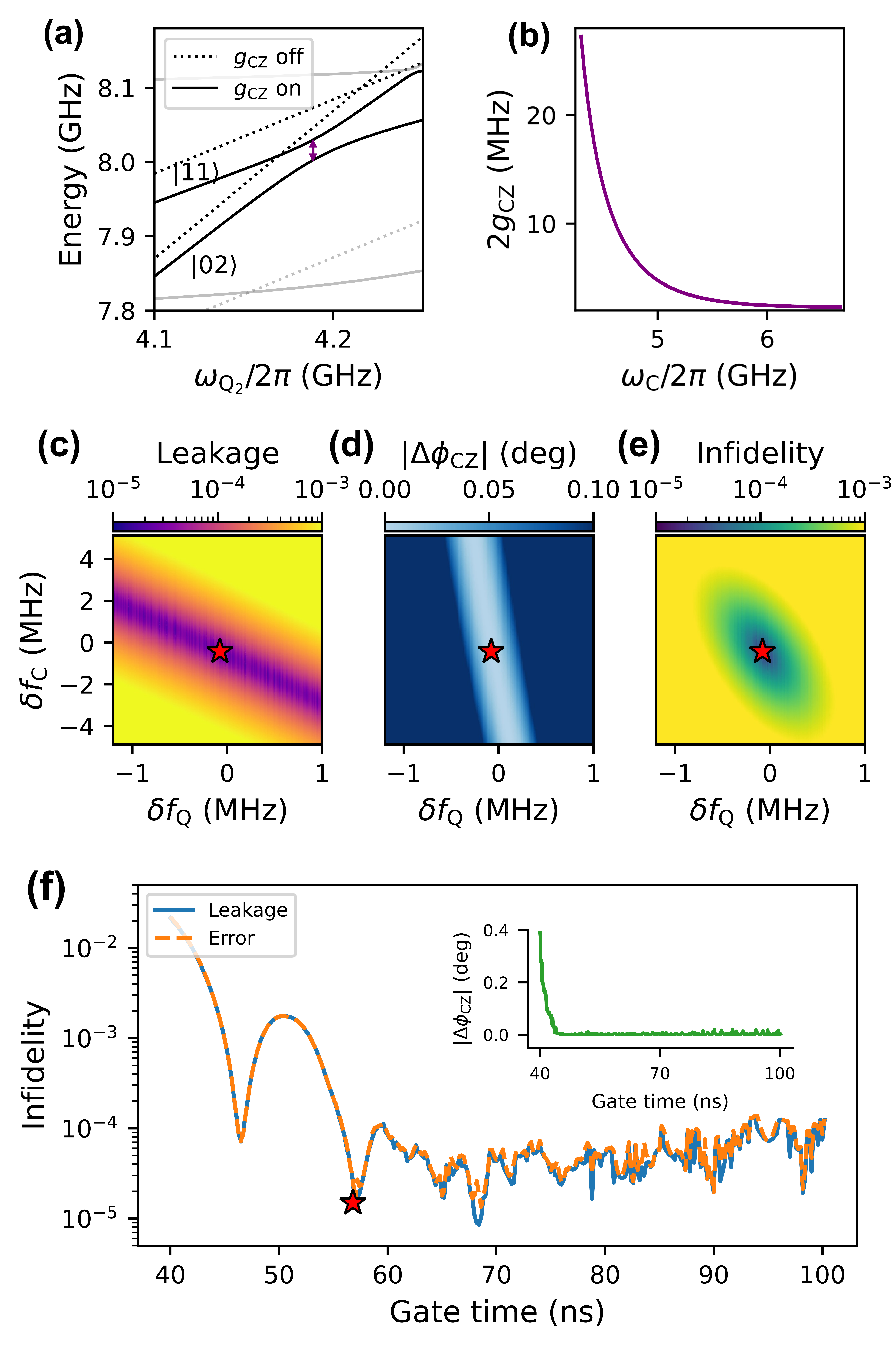}
    \caption{\label{fig:fig3} 
    CZ gate operation and error characterization. (a) Double-excitation spectrum versus $\omega_{\mathrm{Q}_2}$, highlighting the interacting states $\ket{11}$ and $\ket{02}$. Dashed (solid) curves denote coupling off (on); the avoided-crossing splitting (purple arrow) corresponds to the effective coupling $2g_{\mathrm{CZ}}/2\pi$. (b) Extracted $2g_{\mathrm{CZ}}/2\pi$ versus coupler frequency $\omega_{\mathrm{C}}$. (c) Leakage, (d) controlled-phase deviation $|\Delta\phi_{\mathrm{CZ}}|$, and (e) total infidelity from the tune-up at $t_g=58$\:ns, plotted against deviations of the minimum coupler frequency ($\delta f_{\mathrm{C}}$) and qubit detuning ($\delta f_{\mathrm{Q}}$) from the optimal operating point (red star). (f) Leakage (blue) and total infidelity (orange) versus gate time, showing leakage-limited performance (inset: $|\Delta\phi_{\mathrm{CZ}}|$ versus gate time).
}
\end{figure}
% Paragraph 16

\subsection{CZ gate}

We next consider CZ gates implemented by resonantly coupling the $\ket{11}$ state to the second-excited states $\ket{20}$ or $\ket{02}$. Figure~\ref{fig:fig3}(a) shows the double-excitation spectrum, where enabling the couplers produces an avoided crossing between these levels. The resulting splitting defines an effective coupling $2g_{\mathrm{CZ}}/2\pi \approx 25$\:MHz within the dispersive regime, supporting fast coherent oscillations. Completing one oscillation cycle returns the population to $\ket{11}$ while imparting the required controlled phase.

Gate parameters are optimized by adjusting the qubit detuning and the minimum coupler frequency to suppress both leakage and phase errors. 
An example of such tune-up procedure with $t_g=58$\:ns is shown in Fig.~\ref{fig:fig3}(c)–(e). 
Leakage is mainly sensitive to the coupler trajectory amplitude, whereas the implemented controlled phase $\phi_{\mathrm{CZ}}$, ideally $\pi$, exhibits a small deviation $\Delta\phi_{\mathrm{CZ}}=\phi_{\mathrm{CZ}}-\pi$  primarily depending on the qubit detuning. These distinct dependencies allow simultaneous suppression at an optimal operating point.

Figure~\ref{fig:fig3}(f) shows the optimized performance versus gate duration. The near overlap of leakage and total infidelity indicates leakage-dominated behavior with negligible phase contributions. An optimal window around $t_g \sim 58$–$70$\:ns yields coherent errors below $10^{-5}$. Accounting for incoherent errors from finite coherence times, both the iSWAP ($t_g=45$\:ns) and CZ ($t_g=58$\:ns) gates are expected to achieve fidelities of approximately $99.9\:\%$ for $T_1, T_\phi \sim 80\,\mu\text{s}$  (App.~\ref{sec:coherence-2q}).

\begin{figure}[b]
    \centering
    \includegraphics[width=1\linewidth]{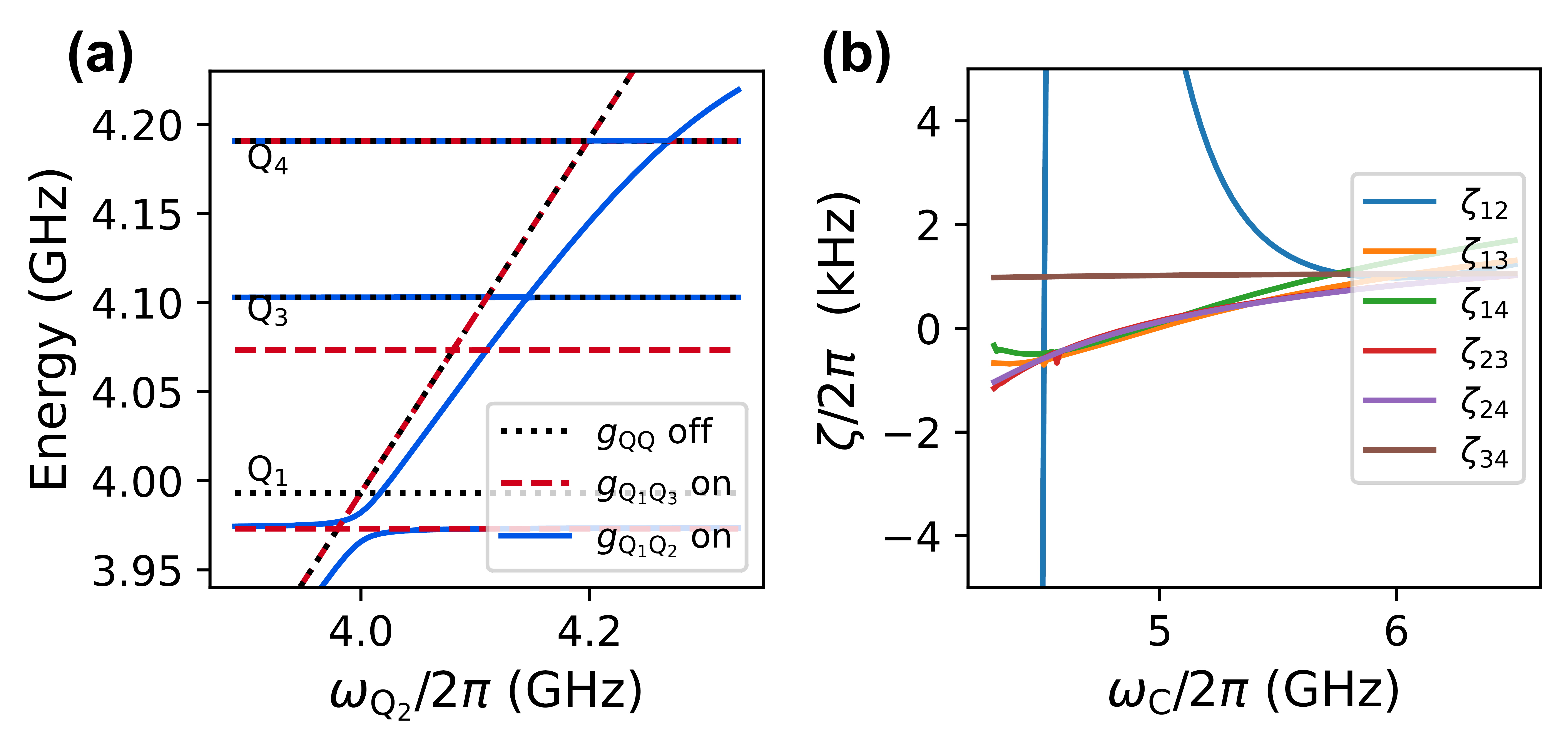}
    \caption{\label{fig:fig4} Selective coupling in a four-qubit system. (a) Single-excitation spectrum obtained by sweeping the frequency of $\mathrm{Q}_2$ under different coupling configurations. An avoided crossing appears only when the $\mathrm{Q}_1$-$\mathrm{Q}_2$ coupling is enabled (solid blue), indicating an effective exchange interaction. No crossings are observed when all couplers are idle (dotted gray) or when only the $\mathrm{Q}_1$-$\mathrm{Q}_3$ pair is coupled (dashed red). (b) Residual $ZZ$ interactions $\zeta_{ij}$ between qubit pairs $(\mathrm{Q}_i, \mathrm{Q}_j)$ as a function of the common coupler frequency $\omega_\mathrm{C}$ of $\mathrm{C_1}$ and $\mathrm{C_2}$, showing suppressed spectator interactions outside the selected pair.
}
\end{figure}
% Paragraph 19
\section{Extension to a four-qubit system}\label{sec:four-qubit}

We extend the architecture to a four-qubit configuration to demonstrate reconfigurable, selective coupling between arbitrary qubit pairs. Four qubits, $\mathrm{Q}_1$--$\mathrm{Q}_4$, are coupled to a common cavity $\mathrm{R}$ through their respective tunable couplers $\mathrm{C}_1$--$\mathrm{C}_4$, forming a shared bus that mediates interactions on demand. Using the same coupling mechanism and gate protocols described above, we simulate selective two-qubit operations such as $\mathrm{CZ}\otimes I \otimes I$ and evaluate their performance in this multi-qubit setting.

% Paragraph 20
\subsection{Selective coupling between qubits}

We first verify selective activation of exchange interactions by examining the single-excitation spectrum. The frequency of $\mathrm{Q}_2$ is swept across those of the other qubits under different coupling configurations, as shown in Fig.~\ref{fig:fig4}(a). With all couplers parked at their idle points (dotted gray), the spectrum exhibits only crossings, indicating negligible residual coupling. The crossings persist when only the $\mathrm{Q}_1$-$\mathrm{Q}_3$ pair is activated (red dashed), where the couplers induce only small dispersive shifts of $\mathrm{Q}_1$ and $\mathrm{Q}_3$. In contrast, enabling $\mathrm{C}_1$ and $\mathrm{C}_2$ (blue solid) produces a clear avoided crossing between $\mathrm{Q}_1$ and $\mathrm{Q}_2$, evidencing an effective exchange interaction. Residual $ZZ$ interactions for all spectator pairs remain below $3$\:kHz [Fig.~\ref{fig:fig4}(b)], confirming that coupling can be activated selectively without introducing appreciable parasitic interactions.

\begin{table}[b] 
\caption{\label{tab:table1} Durations and average fidelities of selective iSWAP gates for all qubit pairs. Device parameters (coupling capacitances and anharmonicities) are identical to those used in Figs.~\ref{fig:fig2}–\ref{fig:fig3}. Fidelity values are rounded to the third decimal place.} \begin{ruledtabular} \begin{tabular}{c c c c c c c} & $\mathrm{Q_1}\text{-}\mathrm{Q_2}$ & $\mathrm{Q_1}\text{-}\mathrm{Q_3}$ & $\mathrm{Q_1}\text{-}\mathrm{Q_4}$ & $\mathrm{Q_2}\text{-}\mathrm{Q_3}$ & $\mathrm{Q_2}\text{-}\mathrm{Q_4}$ & $\mathrm{Q_3}\text{-}\mathrm{Q_4}$ \\ \hline
$\rule{0pt}{2.5ex}t_g$ (ns) & 55.5 & 56.6 & 55.7 & 57.2 & 58.2 & 59.1 \\
$\rule{0pt}{2.5ex}\bar{F}_{L}$ ($\%$) & 99.98 & 99.99 & 99.98 & 99.99 & 99.98 & 99.99 \\ 
$\rule{0pt}{2.5ex}\bar{F}_{P}$ ($\%$) & 99.98 & 99.98 & 99.96 & 99.97 & 99.96 & 99.97 \\ 
$\rule{0pt}{2.5ex}\bar{F}_{ZZ}$ ($\%$) & 99.98 & 99.98 & 99.96 & 99.97 & 99.96 & 99.97 \\ $\rule{0pt}{2.5ex}\bar{F}$ ($\%$) & 99.94 & 99.95 & 99.91 & 99.94 & 99.91 & 99.93 \\
 \end{tabular} \end{ruledtabular} \end{table}
% Paragraph 22

\subsection{Selective two-qubit gates \& error analysis}
To quantify the reliability of reconfigurable couplings, we evaluate two-qubit gate performance for all qubit pairings within the four-qubit system. Gate dynamics are simulated using the time-evolution operator $U$, represented as a $16\times16$ matrix acting on the full Hilbert space. Accordingly, the reported fidelities correspond to the embedded operation on the complete register (e.g., $\mathrm{iSWAP}\otimes I \otimes I$ or $\mathrm{CZ}\otimes I \otimes I$ which are four-qubit processes), and therefore include exchange crosstalk and residual $ZZ$ interactions affecting spectator qubits. Idling frequencies for the qubits and couplers are chosen following the procedure described in App.~\ref{sec:four-qubit-idling}, and gate parameters are optimized using the calibration methods of Sec.~\ref{sec:gate-operations}. To reduce computational cost in the enlarged Hilbert space, we apply the rotating-wave approximation, which renders the Hamiltonian block-diagonal and enables independent simulation of each excitation manifold.

\begin{table}[t] 
\caption{\label{tab:table2} Durations and average fidelities of selective CZ gates for all qubit pairs. Device parameters are identical to those used in Figs.~\ref{fig:fig2}–\ref{fig:fig3}. Fidelity values are rounded to the third decimal place.} \begin{ruledtabular} \begin{tabular}{c c c c c c c} & $\mathrm{Q_1}\text{-}\mathrm{Q_2}$ & $\mathrm{Q_1}\text{-}\mathrm{Q_3}$ & $\mathrm{Q_1}\text{-}\mathrm{Q_4}$ & $\mathrm{Q_2}\text{-}\mathrm{Q_3}$ & $\mathrm{Q_2}\text{-}\mathrm{Q_4}$ & $\mathrm{Q_3}\text{-}\mathrm{Q_4}$ \\ \hline
$\rule{0pt}{2.5ex}t_g$ (ns) & 60.9 & 62.8 & 66.6& 64.2 & 68.2 & 69.0 \\
$\rule{0pt}{2.5ex}\bar{F}_{L}(\%)$ & 99.96& 99.94& 99.93 & 99.96& 99.98& 99.94\\ 
$\rule{0pt}{2.5ex}\bar{F}_{P} (\%)$ & 99.96 & 99.92& 99.92& 99.95& 99.97& 99.92\\ 
$\rule{0pt}{2.5ex}\bar{F}_{ZZ}(\%)$ & 99.96 & 99.92 & 99.92 & 99.95 & 99.97& 99.92 \\
$\rule{0pt}{2.5ex}\bar{F}(\%)$ & 99.95 & 99.92 & 99.91 & 99.95 & 99.95& 99.92 \\
 \end{tabular} \end{ruledtabular} \end{table}

To analyze error mechanisms, we evaluate four fidelities: $\bar{F}_{L}$, $\bar{F}_{P}$, $\bar{F}_{ZZ}$, and $\bar{F}$. 
The first two isolate processes inducing population errors, while the latter two additionally incorporate phase errors arising from residual $ZZ$ interactions and a deviation from the intended controlled phase of the gate.
We define $\bar{F}_{L}=1-\varepsilon_{L}$, where $\varepsilon_{L}$ measures leakage to the non-computational states. We further define $\bar{F}_{P}=1-\varepsilon_{L}-\varepsilon_{PT}$, where $\varepsilon_{PT}$ captures unintended population transfers to the other states in the computational subspace.
For the phase errors, we define $\varepsilon_{ZZ}$ to quantify unwanted phase accumulation arising from residual two-qubit $ZZ$ interactions during the gate, giving $\bar{F}_{ZZ}=1-\varepsilon_{L}-\varepsilon_{PT}-\varepsilon_{ZZ}$. This metric includes crosstalk-induced phases between spectator–spectator and target–spectator pairs, while excluding the intended controlled phase of the active qubit pair. Accordingly, only parasitic phase rotations on states that should ideally remain unchanged contribute to $\varepsilon_{ZZ}$. The total fidelity $\bar{F}$ includes all relative phase deviations and therefore captures the full coherent error of the implemented gate. Details of this simulation are provided in App.~\ref{sec:four-qubit-errors}.

Gate durations and fidelities for selective iSWAP and CZ operations are summarized in Tables~\ref{tab:table1} and \ref{tab:table2}. All qubit pairs exceed $99.9\%$ fidelity, with both unwanted population transfer ($\varepsilon_{PT}$) and residual $ZZ$ phase errors from spectator interactions ($\varepsilon_{ZZ}$) suppressed below $2\times10^{-4}$, indicating minimal spectator crosstalk. For iSWAP gates, performance is primarily limited by residual controlled-phase accumulation; extending the gate duration partially suppresses this effect, leaving an infidelity of $\sim5\times10^{-4}$. In contrast, optimized CZ gates exhibit negligible phase error and are instead leakage-limited (up to $\sim7\times10^{-4}$), with further gains expected from improved pulse shaping and parameter tuning. Overall, the achieved fidelities are comparable to nearest-neighbor architectures with single-transmon tunable couplers, where stray inter-mode couplings ($\sim 0.5-1$ MHz level) similarly constrain $\mathrm{CZ}\otimes I$ errors to the low-$10^{-3}$ range~\cite{coupler-assisted}.

We also estimate gate performance under finite qubit coherence times by incorporating relaxation and dephasing of all four qubits (App.~\ref{sec:coherence-4q}). The full-register fidelity of a CZ gate on the $\mathrm{Q_1}$–$\mathrm{Q_2}$ pair is $\sim99\%$ for $T_1, T_\phi \approx 30\,\mu\text{s}$, improves to $\sim99.8\%$ at $150\,\mu\text{s}$, and exceeds $99.9\%$ near $300\:\mu\text{s}$.

%In contrast, restricting the evaluation to the activated-qubit subspace isolates mainly coherent errors, giving an apparent fidelity of $99.989\%$; including relaxation and dephasing with $T_1, T_\phi = 150\,\mu\text{s}$ still yields fidelities above $99.9\%$. This highlights the importance of full-register evaluation to properly capture leakage and crosstalk in multi-qubit systems.

%We also estimate  performance limited by decoherence, considering the effects of all four qubits (App.~\ref{sec:coherence-4q}). Accounting for decoherence of both target and spectator qubits, the full-register fidelity of a CZ gate acting on the $\mathrm{Q_1}$-$\mathrm{Q_2}$ pair is $\sim99\:\%$ for $T_1, T_\phi \approx 30\,\mu\text{s}$, increases to $\sim99.8\:\%$ for $150\,\mu\text{s}$, and exceeds $99.9\:\%$ for coherence times around $300\,\mu\text{s}$. In contrast, measurements that keep spectator qubits in their ground states probe only the target-qubit subspace and therefore yield higher apparent fidelities (e.g., $99.989\:\%$ for the CZ gate with coherent errors and reaching $99.9\%$ at $150\,\mu\text{s}$). This comparison highlights that full-register evaluation is necessary to properly capture leakage and crosstalk in multi-qubit systems.

\section{Conclusion}

We introduced a cavity-mediated, dynamically reconfigurable coupling architecture that enables fast and selective non-local interactions between superconducting qubits. Numerical simulations show strong effective exchange couplings with excellent idling isolation, supporting fast iSWAP and CZ gates with fidelities exceeding $99.99\%$ and residual $ZZ$ interactions suppressed to the kilohertz level. Selective gates in a four-qubit system further demonstrate flexible connectivity without significantly sacrificing speed or fidelity, allowing flexible qubit placement and efficient implementation of non-local operations.

At the same time, several practical considerations emerge when scaling to larger systems. Because all interactions are mediated by a single shared resonator mode, the bus acts as a common resource, making simultaneous independent gates intrinsically difficult: concurrent couplings compete for the same mode and can mix through unintended multi-qubit pathways rather than forming isolated pairwise interactions. In addition, increasing the number of connected qubits leads to spectral crowding and residual crosstalk, requiring careful frequency planning and time-multiplexed gate scheduling. Finally, the resonator’s physical size and mode spacing ultimately limit how many qubits can be efficiently integrated on a single bus.

These considerations suggest a modular scaling strategy in which moderate-size bus units are interconnected through additional routing or coupling nodes, rather than relying on a single global resonator. Exploring such modular extensions, as well as protocols for coordinating simultaneous multi-qubit operations across shared buses, will be the subject of future work. Within this framework, the proposed architecture provides a practical building block toward scalable, high-connectivity superconducting processors that combine strong, controllable interactions with robust spectator isolation.

\section*{Acknowledgements}
This work was supported by the New Faculty Startup Fund from Seoul National University, POSCO Science Fellowship of POSCO TJ Park Foundation, the National Research Foundation of Korea (NRF) grants supported by the Korea government (MSIT) (No. RS-2024-00334169, No. RS-2024-00413957, No. RS-2024-00467238).

\appendix

\section{Canonical Quantization}\label{sec:quantization}

We consider a set of qubit-coupler units capacitively coupled to a common bus resonator.
The bus is implemented as a coplanar-waveguide (CPW) $\lambda/2$ transmission-line resonator, whose quantization follows the standard distributed treatment based on the telegrapher's equations and normal-mode expansion.
For an open-ended resonator of length $d$ with per-unit-length capacitance $c_0$ and inductance $l_0$, the phase velocity and characteristic impedance are $v_0=1/\sqrt{l_0c_0}$ and $Z_0=\sqrt{l_0/c_0}$, respectively.
Restricting to the fundamental mode, the bus reduces to a single harmonic degree of freedom with flux $\Phi_{\mathrm R}$ and conjugate charge $Q_{\mathrm R}$,
\begin{equation}
H_{\mathrm R}=
\frac{Q_{\mathrm R}^2}{2C_{\mathrm R}}
+\frac{1}{2}C_{\mathrm R}\omega_{\mathrm R}^2\Phi_{\mathrm R}^2,
\label{eq:HR_lumped}
\end{equation}
where $\omega_{\mathrm R}=\pi v_0/d$ is the fundamental angular frequency \cite{BlaisCQED}. 
The effective capacitance associated with the $\lambda/2$ mode is
\begin{equation}
C_{\mathrm R}\equiv \frac{1}{2}\int_0^d c_0\,dx
=\frac{c_0d}{2}
=\frac{\pi}{2\omega_{\mathrm R}Z_0}
=\frac{1}{4f_{\mathrm R}Z_0},
\label{eq:CR_def}
\end{equation}
with $f_{\mathrm R}=\omega_{\mathrm R}/2\pi$.
\begin{figure}[t]
    \centering
    \includegraphics[width=\linewidth]{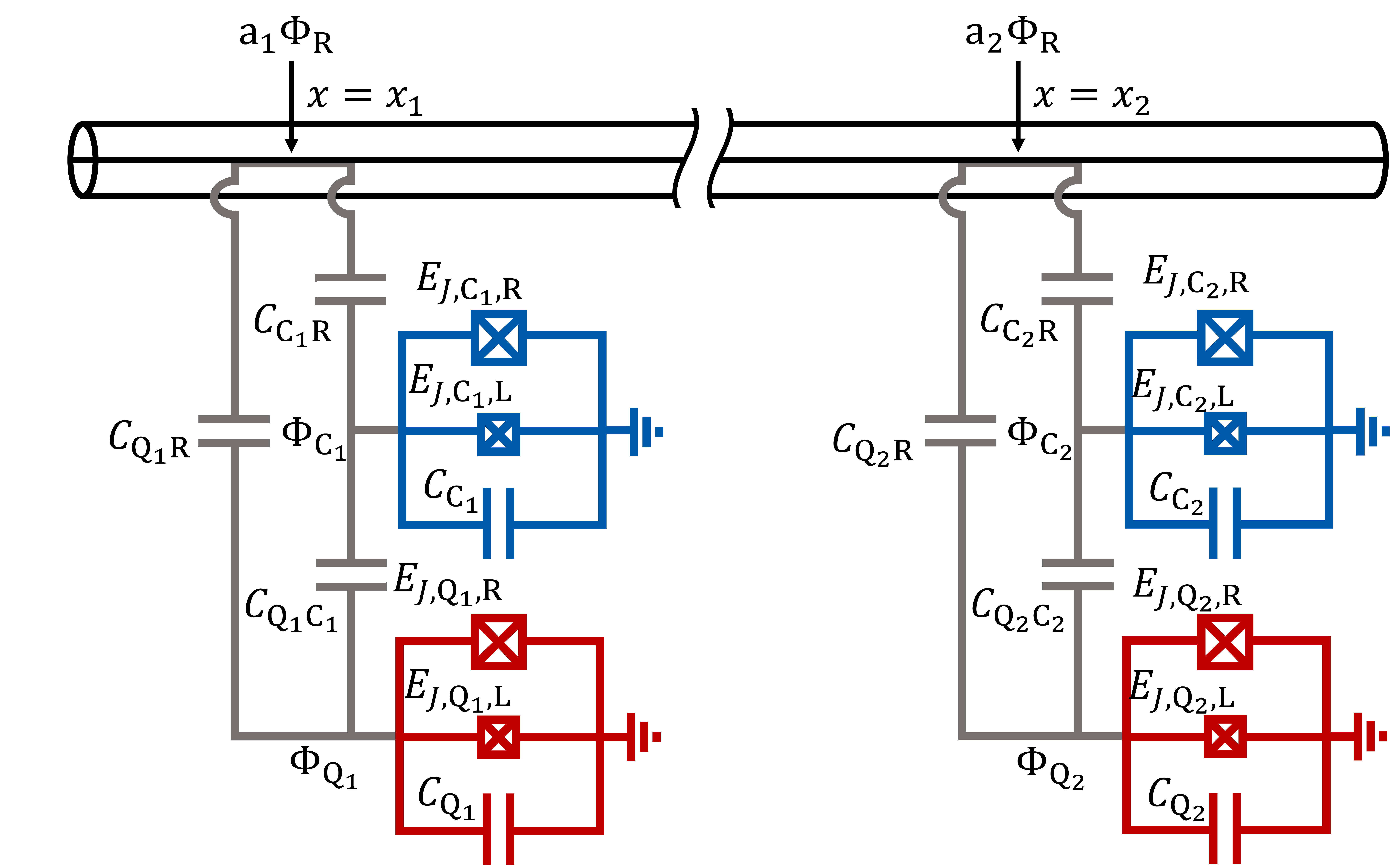}
    \caption{
    Schematic of the circuit architecture used for quantization. A transmission-line resonator serves as a common coupling bus and is restricted to its fundamental mode with flux amplitude $\Phi_{\mathrm{R}}$. Each qubit–coupler unit is capacitively connected to the resonator at position $x_j$, where the local resonator flux is $a_j \Phi_{\mathrm{R}}$, with $a_j$ determined by the spatial profile of the fundamental mode. The coupler (blue) and qubit (red) are modeled as tunable-frequency transmon modes with node flux variables $\Phi_{\mathrm{C}_j}$ and $\Phi_{\mathrm{Q}_j}$, respectively.
    }
    \label{fig:CircuitQuantization}
\end{figure}
The spatial profile of the fundamental resonator mode enters only through the sampled flux at the connection points $x_j$ of the qubit-coupler units. We write
\begin{equation}
\Phi(x_j,t)=a_j\,\Phi_{\mathrm R}(t),
\qquad
a_j\equiv u_1(x_j),
\label{eq:flux_sampling}
\end{equation}
where $u_1(x)$ is the normalized fundamental mode function. With the cosine convention used here,
$u_1(x)=\sqrt{2}\cos(\pi x/d)$ and therefore $a_j=\sqrt{2}\cos(\pi x_j/d)$.

Each qubit and coupler is modeled as a tunable-frequency transmon characterized by a dominant self capacitance $C_{\lambda_k}$ and a nonlinear Josephson element $E_{J,\lambda_k}$, giving a weak anharmonicity. Here, $\lambda \in \{\mathrm{Q, C}\}$ and $k$ denotes the index of each mode.
We choose node fluxes as generalized coordinates: for unit $j$, the qubit and coupler node-flux variables are $\Phi_{\mathrm{Q}_j}$ and $\Phi_{\mathrm{C}_j}$, and the resonator coordinate is $\Phi_{\mathrm R}$.
At the coupling point $x_j$, the resonator contributes the sampled flux $a_j\Phi_{\mathrm R}$.

The circuit Lagrangian is written as $L=T-V$, where the kinetic energy $T$ arises from capacitive elements,
\begin{equation}
\begin{aligned}
T = \frac{1}{2} \Bigg[
& C_{\mathrm{R}} \dot{\Phi}_{\mathrm{R}}^{\,2}
+ \sum_j C_{\mathrm{Q}_j} \dot{\Phi}_{\mathrm{Q}_j}^{\,2}
+ \sum_j C_{\mathrm{C}_j} \dot{\Phi}_{\mathrm{C}_j}^{\,2} \\
& + \sum_j C_{\mathrm{Q}_j\mathrm{R}}
\left( \dot{\Phi}_{\mathrm{Q}_j} - a_j \dot{\Phi}_{\mathrm{R}} \right)^2\\
&+ \sum_j C_{\mathrm{C}_j\mathrm{R}}
\left( \dot{\Phi}_{\mathrm{C}_j} - a_j \dot{\Phi}_{\mathrm{R}} \right)^2 \\
& + \sum_j C_{\mathrm{Q}_j\mathrm{C}_j}
\left( \dot{\Phi}_{\mathrm{Q}_j} - \dot{\Phi}_{\mathrm{C}_j} \right)^2
\Bigg],
\end{aligned}
\label{eq:T_kinetic}
\end{equation}
with $C_{\mathrm{Q}_j}$ and $C_{\mathrm{C}_j}$ the effective self-capacitances of the qubit and coupler, and
$C_{\mathrm{Q}_j\mathrm{R}}$, $C_{\mathrm{C}_j\mathrm{R}}$, and $C_{\mathrm{Q}_j\mathrm{C}_j}$ the coupling capacitances.

The potential energy $V$ is set by the Josephson elements,
\begin{equation}
\begin{aligned}
V &=
\sum_j E_{J,\mathrm{Q}_j}
\left[
1 - \cos\!\left( \frac{2\pi}{\Phi_0} \Phi_{\mathrm{Q}_j} \right)
\right] \\
&\quad+\sum_j E_{J,\mathrm{C}_j}
\left[
1 - \cos\!\left( \frac{2\pi}{\Phi_0} \Phi_{\mathrm{C}_j} \right)
\right],
\end{aligned}
\label{eq:V_potential}
\end{equation}
where $\Phi_0 = h/2e$ is the flux quantum and $E_{J,\mathrm{Q}_j}$ ($E_{J,\mathrm{C}_j}$) denotes the effective Josephson energy of qubit (coupler) $j$.

We now consider a system with two qubit-coupler units, shown in Fig.~\ref{fig:CircuitQuantization}. It is convenient to rewrite $T$ in a compact quadratic form by collecting all node-flux variables into a single vector. We define the generalized flux coordinate vector as
\begin{equation}
\vec{\Phi}=\left(\Phi_{\mathrm{Q}_1},\Phi_{\mathrm{C}_1},\Phi_{\mathrm{R}},\Phi_{\mathrm{C}_2},\Phi_{\mathrm{Q}_2}\right)^{\mathsf{T}}.
\end{equation}
It follows that $T = \frac{1}{2}\, \dot{\vec{\Phi}}^{\,\mathsf{T}} \mathbf{C}\, \dot{\vec{\Phi}}$, where
\begin{widetext}
\begin{equation}
\mathbf{C}=
\resizebox{\columnwidth}{!}{$
\begin{pmatrix}
C_{\mathrm{Q}_1}+C_{\mathrm{Q}_1\mathrm{R}}+C_{\mathrm{Q}_1\mathrm{C}_1}
& -C_{\mathrm{Q}_1\mathrm{C}_1}
& -a_1 C_{\mathrm{Q}_1\mathrm{R}}
& 0
& 0 \\
- C_{\mathrm{Q}_1\mathrm{C}_1}
& C_{\mathrm{C}_1}+C_{\mathrm{C}_1\mathrm{R}}+C_{\mathrm{Q}_1\mathrm{C}_1}
& -a_1 C_{\mathrm{C}_1\mathrm{R}}
& 0
& 0 \\
- a_1 C_{\mathrm{Q}_1\mathrm{R}}
& - a_1 C_{\mathrm{C}_1\mathrm{R}}
& C_{\mathrm{R}} + \sum_{j=1}^{2} a_j \left( C_{\mathrm{Q}_j\mathrm{R}}+C_{\mathrm{C}_j\mathrm{R}} \right)
& - a_2 C_{\mathrm{C}_2\mathrm{R}}
& - a_2 C_{\mathrm{Q}_2\mathrm{R}} \\
0
& 0
& - a_2 C_{\mathrm{C}_2\mathrm{R}}
& C_{\mathrm{C}_2}+C_{\mathrm{C}_2\mathrm{R}}+C_{\mathrm{Q}_2\mathrm{C}_2}
& - C_{\mathrm{Q}_2\mathrm{C}_2} \\
0
& 0
& - a_2 C_{\mathrm{Q}_2\mathrm{R}}
& - C_{\mathrm{Q}_2\mathrm{C}_2}
& C_{\mathrm{Q}_2}+C_{\mathrm{Q}_2\mathrm{R}}+C_{\mathrm{Q}_2\mathrm{C}_2}
\end{pmatrix}
$}
\label{eq:cap_matrix}
\end{equation}
 is a $5\times5$ capacitance matrix determined by the circuit connectivity.
\end{widetext}

The diagonal elements of $\mathbf{C}$ correspond to the total capacitance associated with each node, while the off-diagonal terms encode capacitive couplings between each node. The resonator coordinate $\Phi_{\mathrm{R}}$ couples to each qubit and coupler through the position-dependent coefficients $a_j$.

Given the Lagrangian $L = T - V$, the generalized momenta conjugate to the flux coordinates are defined as
\begin{equation}
Q_{\lambda_k} \equiv \frac{\partial L}{\partial \dot{\Phi}_{\lambda_k}},
\label{eq:gen_mom_def}
\end{equation}
which correspond to the node charges, with $\lambda\in\{\mathrm{Q, C, R}\}$. The index $k$ is removed for $\lambda=\mathrm{R}$. Using the compact representation $T=\tfrac{1}{2}\dot{\vec{\Phi}}^{\,\mathsf{T}}\mathbf{C}\dot{\vec{\Phi}}$,
we obtain the vector relation
\begin{equation}
\vec{Q} = \mathbf{C}\,\dot{\vec{\Phi}},
\label{eq:q_equals_Cphidot}
\end{equation}
where $\vec{Q}=(Q_{\mathrm{Q}_1},Q_{\mathrm{C}_1},Q_{\mathrm{R}},Q_{\mathrm{C}_2},Q_{\mathrm{Q}_2})^{\mathsf{T}}$. The classical Hamiltonian then follows from the Legendre transform,
\begin{equation}
H = \sum_{\lambda,k} Q_{\lambda_k} \dot{\Phi}_{\lambda_k} - L
= \frac{1}{2}\,\vec{Q}^{\,\mathsf{T}}\,\mathbf{C}^{-1}\vec{Q} + V,
\label{eq:H_classical}
\end{equation}
where the inverse capacitance matrix $\mathbf{C}^{-1}$ is given by
\begin{equation}
\mathbf{C}^{-1} \approx
\begin{pmatrix}
D_{11} & D_{12} & D_{13} & D_{14} & D_{15} \\
D_{21} & D_{22} & D_{23} & D_{24} & D_{25} \\
D_{31} & D_{32} & D_{33} & D_{34} & D_{35} \\
D_{41} & D_{42} & D_{43} & D_{44} & D_{45} \\
D_{51} & D_{52} & D_{53} & D_{54} & D_{55} \\
\end{pmatrix}
\label{eq:cap_matrix_inverse}
\end{equation}
with
\begin{equation}
\begin{aligned}
& D_{11} \approx \dfrac{1}{C_{\mathrm{Q}_1}}, \quad
D_{22} \approx \dfrac{1}{C_{\mathrm{C}_1}}, \quad
D_{33} \approx \dfrac{1}{C_{\mathrm{R}}}, \\
& D_{44} \approx \dfrac{1}{C_{\mathrm{C}_2}}, \quad
D_{55} \approx \dfrac{1}{C_{\mathrm{Q}_2}},
\end{aligned}
\end{equation}
\begin{equation}
\begin{aligned}
& D_{12}=D_{21} \approx \dfrac{C_{\mathrm{Q}_1\mathrm{C}_1}}{C_{\mathrm{Q}_1}C_{\mathrm{C}_1}}, \quad
D_{23}=D_{32} \approx a_1\dfrac{C_{\mathrm{C}_1\mathrm{R}}}{C_{\mathrm{C}_1}C_{\mathrm{R}}}, \\
& D_{34}=D_{43} \approx a_2\dfrac{C_{\mathrm{C}_2\mathrm{R}}}{C_{\mathrm{C}_2}C_{\mathrm{R}}}, \quad
D_{45}=D_{54} \approx \dfrac{C_{\mathrm{Q}_2\mathrm{C}_2}}{C_{\mathrm{Q}_2}C_{\mathrm{C}_2}},
\end{aligned}
\end{equation}
\begin{equation}
\begin{aligned}
& D_{13}=D_{31} \approx a_1\dfrac{C_{\mathrm{Q}_1\mathrm{R}}C_{\mathrm{C}_1}+C_{\mathrm{Q}_1\mathrm{C}_1}C_{\mathrm{C}_1\mathrm{R}}}{C_{\mathrm{Q}_1}C_{\mathrm{C}_1}C_{\mathrm{R}}}, \\
& D_{24}=D_{42} \approx a_1a_2\dfrac{C_{\mathrm{C}_1\mathrm{R}}C_{\mathrm{C}_2\mathrm{R}}}{C_{\mathrm{C}_1}C_{\mathrm{C}_2}C_{\mathrm{R}}}, \\
& D_{35}=D_{53} \approx a_2\dfrac{C_{\mathrm{Q}_2\mathrm{R}}C_{\mathrm{C}_2}+C_{\mathrm{Q}_2\mathrm{C}_2}C_{\mathrm{C}_2\mathrm{R}}}{C_{\mathrm{Q}_2}C_{\mathrm{C}_2}C_{\mathrm{R}}},
\end{aligned}
\end{equation}
\begin{equation}
\begin{aligned}
D_{14}=D_{41} &\approx a_1 a_2
\frac{C_{\mathrm{C}_2\mathrm{R}}\left(C_{\mathrm{Q}_1\mathrm{R}}C_{\mathrm{C}_1}+C_{\mathrm{Q}_1\mathrm{C}_1}C_{\mathrm{C}_1\mathrm{R}}\right)}
     {C_{\mathrm{Q}_1}C_{\mathrm{C}_1}C_{\mathrm{C}_2}C_{\mathrm{R}}}, \\
D_{25}=D_{52} &\approx a_1 a_2
\frac{C_{\mathrm{C}_1\mathrm{R}}\left(C_{\mathrm{Q}_2\mathrm{R}}C_{\mathrm{C}_2}+C_{\mathrm{Q}_2\mathrm{C}_2}C_{\mathrm{C}_2\mathrm{R}}\right)}
     {C_{\mathrm{Q}_2}C_{\mathrm{C}_1}C_{\mathrm{C}_2}C_{\mathrm{R}}}
\end{aligned}
\end{equation}
\begin{align}
D_{15}&=D_{51} \nonumber\\
&\approx \frac{a_1 a_2}{C_\mathrm{R}}
\frac{
\left(C_{\mathrm{Q}_1\mathrm{R}}C_{\mathrm{C}_1}+C_{\mathrm{Q}_1\mathrm{C}_1}C_{\mathrm{C}_1\mathrm{R}}\right)
}
{C_{\mathrm{Q}_1}C_{\mathrm{C}_1}}\nonumber\\
&\qquad\times \frac{
\left(C_{\mathrm{Q}_2\mathrm{R}}C_{\mathrm{C}_2}+C_{\mathrm{Q}_2\mathrm{C}_2}C_{\mathrm{C}_2\mathrm{R}}\right)}{C_{\mathrm{Q}_2}C_{\mathrm{C}_2}}
\end{align}

With $C_{\lambda_k}\gg C_{\mathrm{Q}_j\mathrm{C}_j}, C_{\mathrm{C}_j\mathrm{R}}\gg C_{\mathrm{Q}_j\mathrm{R}}$, we retain the leading-order terms.  We find $D_{14}=D_{41}\approx 0$, $D_{25}=D_{52}\approx 0$, and $D_{15}=D_{51}\approx 0$. 
The terms in $D_{13}=D_{31}$ and $D_{35}=D_{53}$, which correspond to direct couplings between each qubit and the resonator, remain finite.
The element $D_{24}=D_{42}$, representing a direct coupling between the two couplers scales as $D_{24}\sim a_1 a_2\, C_{\mathrm{C}_1\mathrm{R}}C_{\mathrm{C}_2\mathrm{R}}/(C_{\mathrm{C}_1}C_{\mathrm{C}_2}C_{\mathrm{R}})$. Despite the large $C_{\mathrm{R}}$, we keep this element because it is amplified by the product of the large $C_{\mathrm{C}_j\mathrm{R}}$ capacitances in the numerator. Elements in the main diagonal correspond to the renormalized self capacitances and the first off-diagonal elements are the nearest-neighbor couplings.

Canonical quantization proceeds by promoting fluxes and charges to operators satisfying
\begin{equation}
[\Phi_\lambda, Q_{\lambda'}] = i\hbar\,\delta_{\lambda\lambda'} .
\end{equation}
For each transmon mode $\lambda$, we work in the transmon regime $E_{J,\lambda}\gg E_{C_\lambda}$ with $E_{C_\lambda}=e^2/(2C_\lambda)$.
Expanding the cosine to quartic order yields a weakly anharmonic oscillator. Introducing bosonic operators,
\begin{align}
\Phi_\lambda &= \Phi_{\mathrm{zpf},\lambda}\left(a_\lambda + a_\lambda^\dagger\right), \\
Q_\lambda &= i Q_{\mathrm{zpf},\lambda}\left(a_\lambda^\dagger - a_\lambda\right),
\end{align}
with zero-point fluctuations
\begin{equation}
\Phi_{\mathrm{zpf},\lambda}
= \sqrt{\frac{\hbar}{2C_\lambda \omega_\lambda}},
\qquad
Q_{\mathrm{zpf},\lambda}
= \sqrt{\frac{\hbar C_\lambda \omega_\lambda}{2}} .
\end{equation}
The mode frequencies and anharmonicities are
\begin{equation}
\omega_\lambda = \sqrt{8E_{J\lambda}E_{C\lambda}} - E_{C\lambda},
\qquad
\alpha_\lambda = -E_{C\lambda},
\end{equation}
with $\alpha_{\mathrm{R}}=0$ for the harmonic resonator mode.

\subsubsection*{Self Hamiltonians}

The self Hamiltonian of each mode $\lambda \in
\{\mathrm{Q}_1,\mathrm{C}_1,\mathrm{R},\mathrm{C}_2,\mathrm{Q}_2\}$ takes the Duffing form
\begin{equation}
H_\lambda
=
\omega_\lambda a_\lambda^\dagger a_\lambda
+
\frac{\alpha_\lambda}{2}
a_\lambda^\dagger a_\lambda^\dagger a_\lambda a_\lambda .
\end{equation}

\subsubsection*{Interaction Hamiltonians}

Using the leading-order elements of the inverse capacitance matrix,
the interaction Hamiltonian can be written as a sum of bilinear couplings.

\paragraph*{Qubit-coupler coupling.}
For each qubit-coupler pair,
\begin{equation}
H_{\mathrm{Q}_j\mathrm{C}_j}
=
-g_{\mathrm{Q}_j\mathrm{C}_j}
\left(
a_{\mathrm{Q}_j}^\dagger-a_{\mathrm{Q}_j}\right)
\left(a_{\mathrm{C}_j}^\dagger-a_{\mathrm{C}_j}\right),
\quad j\in\{1,2\},
\end{equation}
with coupling strength
\begin{equation}
g_{\mathrm{Q}_j\mathrm{C}_j}
=
\frac{1}{2}
\frac{C_{\mathrm{Q}_j\mathrm{C}_j}}{\sqrt{C_{\mathrm{Q}_j}C_{\mathrm{C}_j}}}
\sqrt{\omega_{\mathrm{Q}_j}\omega_{\mathrm{C}_j}} .
\end{equation}

\paragraph*{Coupler-resonator coupling.}
The coupling between each coupler and the bus resonator is
\begin{equation}
H_{\mathrm{C}_j\mathrm{R}}
=
-g_{\mathrm{C}_j\mathrm{R}}
\left(
a_{\mathrm{C}_j}^\dagger-a_{\mathrm{C}_j}\right)
\left(a_{\mathrm{R}}^\dagger-a_{\mathrm{R}}\right),
\qquad j\in\{1,2\},
\end{equation}
where
\begin{equation}
g_{\mathrm{C}_j\mathrm{R}}
=
\frac{a_j}{2}
\frac{C_{\mathrm{C}_j\mathrm{R}}}{\sqrt{C_{\mathrm{C}_j}C_{\mathrm{R}}}}
\sqrt{\omega_{\mathrm{C}_j}\omega_{\mathrm{R}}} .
\end{equation}

\paragraph*{Direct qubit-resonator coupling.}
A small direct qubit-resonator interaction remains,
\begin{equation}
H_{\mathrm{Q}_j\mathrm{R}}
=
-g_{\mathrm{Q}_j\mathrm{R}}
\left(
a_{\mathrm{Q}_j}^\dagger-a_{\mathrm{Q}_j}\right)
\left(a_{\mathrm{R}}^\dagger-a_{\mathrm{R}}\right),
\qquad j\in\{1,2\},
\end{equation}
with
\begin{equation}
g_{\mathrm{Q}_1\mathrm{R}}
=
\frac{a_1}{2}
\frac{C_{\mathrm{Q}_1\mathrm{R}}C_{\mathrm{C}_1}+C_{\mathrm{Q}_1\mathrm{C}_1}C_{\mathrm{C}_1\mathrm{R}}}{C_{\mathrm{Q}_1}C_{\mathrm{C}_1}}
\frac{1}{\sqrt{C_{\mathrm{Q}_1}C_{\mathrm{R}}}}
\sqrt{\omega_{\mathrm{Q}_1}\omega_{\mathrm{R}}} ,
\end{equation}
and
\begin{equation}
g_{\mathrm{Q}_2\mathrm{R}}
=
\frac{a_2}{2}
\frac{C_{\mathrm{Q}_2\mathrm{R}}C_{\mathrm{C}_2}+C_{\mathrm{Q}_2\mathrm{C}_2}C_{\mathrm{C}_2\mathrm{R}}}{C_{\mathrm{Q}_2}C_{\mathrm{C}_2}}
\frac{1}{\sqrt{C_{\mathrm{Q}_2}C_{\mathrm{R}}}}
\sqrt{\omega_{\mathrm{Q}_2}\omega_{\mathrm{R}}} .
\end{equation}

\paragraph*{Coupler-coupler coupling.}
The only remaining cross-block element of $\mathbf{C}^{-1}$ generates a
coupler-coupler interaction,
\begin{equation}
H_{\mathrm{C}_1\mathrm{C}_2}
=
-g_{\mathrm{C}_1\mathrm{C}_2}
\left(
a_{\mathrm{C}_1}^\dagger-a_{\mathrm{C}_1}\right)
\left(a_{\mathrm{C}_2}^\dagger-a_{\mathrm{C}_2}\right),
\end{equation}
where
\begin{equation}
g_{\mathrm{C}_1\mathrm{C}_2}
=
\frac{a_1 a_2}{2}
\frac{C_{\mathrm{C}_1\mathrm{R}}C_{\mathrm{C}_2\mathrm{R}}}
{\sqrt{C_{\mathrm{C}_1}C_{\mathrm{C}_2}}\,C_{\mathrm{R}}}
\sqrt{\omega_{\mathrm{C}_1}\omega_{\mathrm{C}_2}}.
\end{equation}

\subsubsection*{Total Hamiltonian}

Collecting all the terms, the quantum mechanical Hamiltonian of the system is
\begin{equation}
H
=
\sum_{\lambda}
H_\lambda
+
\sum_{j\in\{1,2\}}
\left(
H_{\mathrm{Q}_j\mathrm{C}_j}
+
H_{\mathrm{C}_j\mathrm{R}}
+
H_{\mathrm{Q}_j\mathrm{R}}
\right)
+
H_{\mathrm{C}_1\mathrm{C}_2},
\end{equation}
which forms the starting point for deriving the effective cavity-mediated qubit-qubit interaction. For notational simplicity, the derivation above was written explicitly for two qubit-coupler units ($j=1,2$). 

The circuit quantization procedure generalizes to an arbitrary number $N$ of qubit-coupler units capacitively coupled to the bus resonator. The total Hamiltonian is written as following:

\begin{equation}
H = H_{\mathrm{self}} + H_{\mathrm{int}},
\end{equation}
where
\begin{equation}
H_{\mathrm{self}} =
\sum_{m,i}
\left(
\omega_{m_i} a_{m_i}^\dagger a_{m_i}
+ \frac{\alpha_{m_i}}{2}
a_{m_i}^\dagger a_{m_i}^\dagger a_{m_i} a_{m_i}
\right),
\label{eq:self-H-appendix}
\end{equation}
and
\begin{eqnarray}
H_{\mathrm{int}}
&=&
-\sum_i \sum_{\substack{m,n \\ m \neq n}}
g_{m_i n_i}
\left(a_{m_i}^\dagger - a_{m_i}\right)
\left(a_{n_i}^\dagger - a_{n_i}\right)
\nonumber \\
&&
-\sum_{\substack{i,j \\ i>j}}
g_{\mathrm{C}_i \mathrm{C}_j}
\left(a_{\mathrm{C}_i}^\dagger - a_{\mathrm{C}_i}\right)
\left(a_{\mathrm{C}_j}^\dagger - a_{\mathrm{C}_j}\right)    .
\label{eq:int-H-appendix}
\end{eqnarray}

In Eqs.~\eqref{eq:self-H-appendix} and \eqref{eq:int-H-appendix}, the index $i$ labels individual qubit-coupler units coupled to the common bus resonator, while $m$ and $n$ denote the internal modes within each unit. Specifically, $m,n\in\{\mathrm{Q},\mathrm{C},\mathrm{R}\}$ correspond to the qubit, coupler, and resonator degrees of freedom, respectively. The operator $a_{m_i}$ annihilates an excitation in mode $m$ of unit $i$, with corresponding mode frequency $\omega_{m_i}$ and anharmonicity $\alpha_{m_i}$. The resonator mode is harmonic, for which $\alpha_{\mathrm R}=0$.

The self Hamiltonian $H_{\mathrm{self}}$ collects all local contributions, describing independent weakly anharmonic transmon modes for the qubits and couplers together with a single harmonic resonator mode. The interaction Hamiltonian $H_{\mathrm{int}}$ contains bilinear exchange terms arising from capacitive couplings. The first summation in Eq.~\eqref{eq:int-H-appendix} accounts for interactions between distinct modes within the same qubit-coupler unit, whereas the second summation describes inter-unit coupler-coupler interactions.

Note that in numerical simulations done in this paper, we assume uniform coupling with $C_{\mathrm{Q}_j\mathrm{C}_j}=C_{\mathrm{QC}}$,
$C_{\mathrm{C}_j\mathrm{R}}=C_{\mathrm{CR}}$,
$C_{\mathrm{Q}_j\mathrm{R}}=C_{\mathrm{QR}}$, and $a_j=1$. For different values of $a_j$, uniform coupling between each qubit–coupler unit and the resonator can be achieved by tuning the corresponding capacitances.

\section{Frequency configuration}\label{sec:frequency-config}
In this section, we outline the considerations used to determine the mode frequencies shown in Fig.~\ref{fig:fig1}(c), with the goal of achieving desired tunable couplings between qubit-qubit and qubit-resonator pairs. 

We first examine the effective qubit–resonator coupling,
\begin{equation}
\tilde{g}_{\mathrm{Q}_i\mathrm{R}}
=
\frac{
g_{\mathrm{Q}_i\mathrm{C}_i}
g_{\mathrm{C}_i\mathrm{R}}
}{\Delta_i}
+
g_{\mathrm{Q}_i\mathrm{R}},
\end{equation}
where $\Delta_{{m_i}{n_j}}=\omega_{m_i}-\omega_{n_j}$ for $m,n\in\{\mathrm{Q,C,R}\}$ and
\[
\Delta_i^{-1}
=
\frac{1}{2}
\left(
\Delta_{\mathrm{Q}_i\mathrm{C}_i}^{-1}
+
\Delta_{\mathrm{R}\mathrm{C}_i}^{-1}
\right).
\]
Because the capacitive couplings $g_{\mathrm{Q}_i\mathrm{C}_i}$ and $g_{\mathrm{C}_i\mathrm{R}}$ are positive, destructive interference between the indirect and direct paths requires $\Delta_i<0$ to achieve $\tilde{g}_{\mathrm{Q}_i\mathrm{R}}\approx0$. This condition defines the idling configuration in which each qubit is effectively decoupled from the resonator.

When $\tilde{g}_{\mathrm{Q}_i\mathrm{R}}$ is tuned away from zero by adjusting the coupler frequency, interactions are mediated through the resonator. The effective Hamiltonian takes the form
\begin{align}
\tilde{H}_{\mathrm{self}} &=
\sum_i
\left(
\tilde{\omega}_{\mathrm{Q}_i}
a_{\mathrm{Q}_i}^\dagger a_{\mathrm{Q}_i}
+
\frac{\alpha_{\mathrm{Q}_i}}{2}
a_{\mathrm{Q}_i}^{\dagger 2}
a_{\mathrm{Q}_i}^2
\right)
+
\tilde{\omega}_{\mathrm{R}}
a_{\mathrm{R}}^\dagger a_{\mathrm{R}}, \\
\tilde{H}_{\mathrm{int}} &=
\sum_i
\tilde{g}_{\mathrm{Q}_i\mathrm{R}}
\left(
a_{\mathrm{Q}_i}^\dagger a_{\mathrm{R}}
+
a_{\mathrm{Q}_i} a_{\mathrm{R}}^\dagger
\right).
\end{align}
For two activated qubits $i$ and $j$, the dispersive interaction yields an effective exchange coupling
\[
\tilde{g}_{\mathrm{Q}_i\mathrm{Q}_j}
\approx
\frac{1}{2}
\tilde{g}_{\mathrm{Q}_i\mathrm{R}}
\tilde{g}_{\mathrm{Q}_j\mathrm{R}}
\left(
\tilde{\Delta}_{\mathrm{Q}_i\mathrm{R}}^{-1}
+
\tilde{\Delta}_{\mathrm{Q}_j\mathrm{R}}^{-1}
\right),
\]
whose magnitude increases as the qubit–resonator detunings decrease. To enhance this interaction while maintaining the idle cancellation condition $\Delta_i<0$, the couplers are biased at frequencies higher than both the qubits and the resonator.

Strong and fast gates further benefit from larger coupler–resonator coupling $g_{\mathrm{CR}}$ relative to the qubit–coupler coupling $g_{\mathrm{QC}}$. This hierarchy promotes stronger hybridization between the couplers and the resonator, thereby enhancing the mediated qubit–qubit interaction while avoiding excessive direct qubit–coupler mixing. With the couplers initially parked at high frequencies, interactions are activated simply by lowering the selected coupler frequencies toward the qubit–resonator band.

Finally, we consider the relative detuning between the qubits and the resonator. Lowering the coupler frequencies induces downward dispersive shifts of both the qubit and resonator modes. Because the resonator couples to multiple couplers, its frequency shift is larger than that of each qubit. To avoid accidental resonances or crossings during activation, we therefore place the qubit operating frequencies above the resonator. This configuration also reduces resonator population during gate operation, mitigating relaxation losses associated with the bus mode. Combining these considerations, we adopt the frequency arrangement shown in Fig.~\ref{fig:fig1}(c).

\begin{figure*}[t]  % figure* → 두 컬럼 전체 폭 사용
    \centering
    \includegraphics[width=\textwidth]{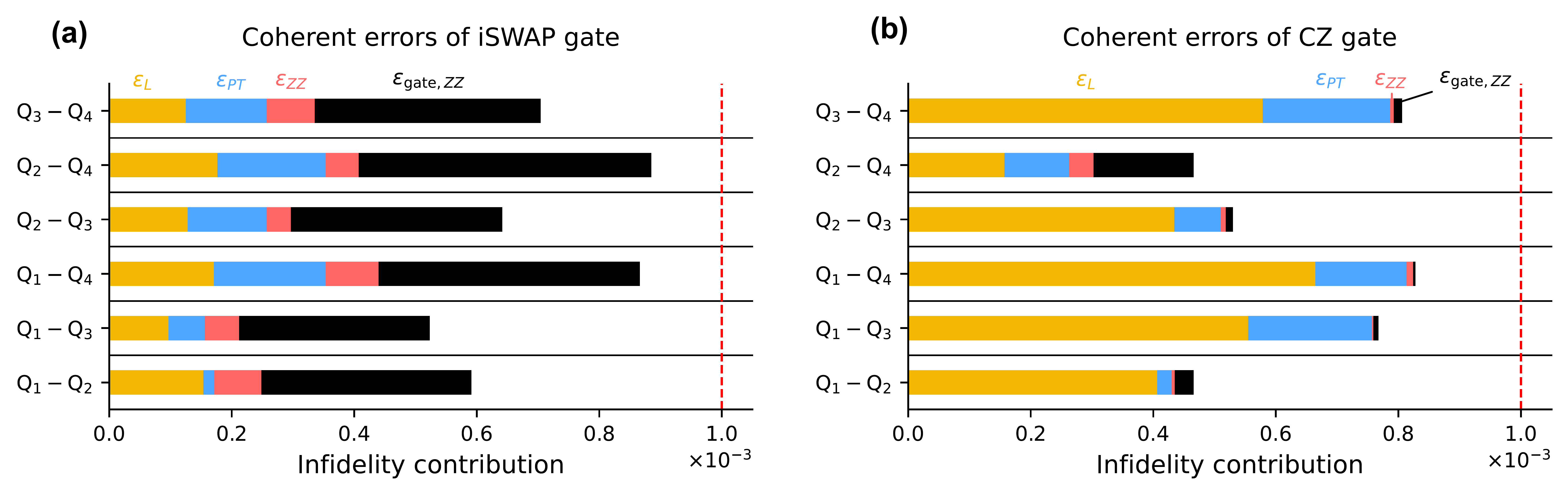}
    \caption{Breakdown of coherent error contributions for (a) selective iSWAP gates and (b) CZ gates summarized in Tables~\ref{tab:table1}-\ref{tab:table2}. 
The plotted components are $\varepsilon_{L}$ (leakage to the outside of the computational subspace), 
$\varepsilon_{PT}$ (errors from unintended population transfer within the computational subspace), 
$\varepsilon_{ZZ}$ (spectator-induced phase errors from residual $ZZ$ interactions), and 
$\varepsilon_{\mathrm{gate},ZZ}$ (phase errors associated with the intended controlled phase of the target qubit pair).}\label{fig:errorbudget}
\end{figure*}

\section{Average gate fidelity}\label{sec:gate-fidelity}
The average gate fidelity is evaluated by reconstructing the operator
$U$ of the simulated gate from numerical time evolution and comparing it to the
ideal target unitary $U_{\mathrm{id}}$.
The fidelity is computed using
\begin{equation}
\bar{F}
=
\frac{|\mathrm{tr}(U_{\mathrm{id}}^{\dagger}U)|^2
+
\mathrm{tr}(U^{\dagger}U)}
{d(d+1)},
\end{equation}
where $d=2^n$ is the Hilbert-space dimension of an $n$-qubit system \cite{dtcfidel}. For reference, $U$ is a $4\times4$ matrix for two-qubit simulations and a
$16\times16$ matrix for four-qubit simulations.

To obtain $U$, we simulate the evolution of each computational basis state
$\ket{\psi_{i_1\cdots i_n}}$ and assemble the matrix elements as
\begin{equation}
U_{ij} = \langle \psi_i \mid \psi'_j \rangle ,
\end{equation}
where $\ket{\psi'_j}$ is the final state resulting from time-evolution of an initial basis state $\ket{\psi_j}$.
Single-qubit phases are removed from each column to isolate relative phases
relevant to the two-qubit gate operation.

To separate errors related to populations, such as leakage from phase errors, we additionally define a population-only
fidelity $\bar{F}_P$ by replacing both $U$ and $U_{\mathrm{id}}$ with their
element-wise magnitudes,
\begin{equation}
U'_{ij}=|U_{ij}|, \qquad
U'_{\mathrm{id},ij}=|U_{\mathrm{id},ij}|,
\end{equation}
thereby discarding phase information.
The population error is then $1-\bar{F}_P$, while the remaining difference
$\bar{F}-\bar{F}_P$ quantifies phase errors accumulated during the gate.

\section{Simulation details of two-qubit gates in a four-qubit system}
\label{sec:four-qubit-simulation}

\subsection{Determination of idling frequencies}
\label{sec:four-qubit-idling}

For multi-qubit operation, we choose an idling configuration that avoids frequency
collisions and suppresses spectator-induced leakage during gate execution.
We first assign the bare qubit frequencies to avoid near-degeneracies such as
$\omega_{\mathrm Q_i}\approx\omega_{\mathrm Q_j}$,
$\omega_{\mathrm Q_i}\approx\omega_{\mathrm Q_j}+\alpha_{\mathrm Q}$,
and two-photon conditions
$\omega_{\mathrm Q_i}+\omega_{\mathrm Q_j}\approx 2\omega_{\mathrm Q_k}+\alpha_{\mathrm Q}$.
Gate operations are performed with the participating qubits biased near the bus
frequency [Fig.~\ref{fig:fig1}(c)] to enhance cavity-mediated interactions.

We then determine the idle frequency of each coupler $\omega_{\mathrm C_k}$ by
numerical optimization. For each candidate $\omega_{\mathrm C_k}$, we simulate
representative two-qubit gates on other qubit pairs and monitor leakage from
computational states that include an excitation in $\mathrm Q_k$.
As an example, to set the idle point of $\mathrm C_3$, we simulate an iSWAP gate on
the $\mathrm Q_1$-$\mathrm Q_2$ pair and evaluate leakage from the state $\ket{1110}$; the
value of $\omega_{\mathrm C_3}$ minimizing this leakage is chosen.
This procedure yields coupler idle points that keep spectator qubits effectively
decoupled during operations on arbitrary target pairs.

\subsection{Calculation and analysis of coherent errors} \label{sec:four-qubit-errors}

In the four-qubit system, coherent errors arise not only from non-adiabatic leakage
into the couplers/resonator but also from unintended exchange and residual $ZZ$
interactions involving spectator qubits. To quantify these effects in a way that is
diagnostic for the cavity-mediated, reconfigurable architecture, we decompose the total
coherent error into four contributions:
(i) $\varepsilon_{L}$, leakage to the non-computational states;
(ii) $\varepsilon_{PT}$, population loss from unwanted exchanges between the other computational basis states;  
(iii) $\varepsilon_{ZZ}$, parasitic phase accumulation caused by residual $ZZ$
interactions involving at least one spectator qubit (spectator-spectator and
spectator-target pairs); and
(iv) $\varepsilon_{\mathrm{gate},ZZ}$, the phase error associated with the target
qubit pair itself (i.e., deviation from the intended controlled phase of the gate).

Following Sec.~\ref{sec:four-qubit}, we extract these errors using four fidelities,
$\bar{F}_{L}$, $\bar{F}_{P}$, $\bar{F}_{ZZ}$, and $\bar{F}$, defined such that
\begin{align*}
    \bar{F}_{L}&=1-\varepsilon_{L},\\
    \bar{F}_{P}&=1 -\varepsilon_{L} - \varepsilon_{PT},\\
    \bar{F}_{ZZ}&=1-\varepsilon_{L} - \varepsilon_{PT} - \varepsilon_{ZZ},\\
    \bar{F}&=1-\varepsilon_{L} - \varepsilon_{PT} - \varepsilon_{ZZ}-\varepsilon_{\mathrm{gate},ZZ}.
\end{align*}
Here, $\varepsilon_{L}$ and $\varepsilon_{\mathrm{gate},ZZ}$ are primarily set by
the gate trajectory design (adiabaticity and target-pair phase engineering), whereas
$\varepsilon_{PT}$ and $\varepsilon_{ZZ}$ directly quantify parasitic interactions
with spectator qubits.

\paragraph*{Population metrics.}
We reconstruct the gate matrix $U$ from numerical time evolution as described in
App.~\ref{sec:gate-fidelity}. Note that $U_{id}$, the ideal unitary matrix of the desired operation now takes a form such as $CZ\otimes I\otimes I$. To isolate population-related errors, we evaluate fidelities using the
phase-stripped matrices $U'_{ij}=|U_{ij}|$ and $U'_{\mathrm{id},ij}=|U_{\mathrm{id},ij}|$,
which yields $\bar{F}_{P}$ and therefore the total population error $1-\bar{F}_P$.
To separate leakage out of the computational subspace, we construct an auxiliary
matrix $U''$ that has the same single-nonzero-entry structure as $U_{\mathrm{id}}$:
for each column $j$, the nonzero entry is placed at the ideal output row and set to
\begin{equation}
u''_j=\left(\sum_{i\in \mathcal{C}} |U_{ij}|^2\right)^{1/2},
\end{equation}
where the sum runs over the computational subspace denoted as $\mathcal{C}$.
Thus $u''_j$ is the total amplitude remaining inside $\mathcal{C}$,
independent of how that population is redistributed among computational states.
The fidelity computed using $U''$ only considers leakage, therefore corresponds to $\bar{F}_{L}$ and defines
$\varepsilon_{L}=1-\bar{F}_{L}$; the remaining difference
$\varepsilon_{PT}=\bar{F}_{L}-\bar{F}_{P}$ quantifies unintended population transfer
within the computational manifold.

\paragraph*{Phase metrics and separation of $\varepsilon_{ZZ}$ vs.\ $\varepsilon_{\mathrm{gate},ZZ}$.}
To evaluate phase errors, we remove single-qubit phases and work with relative phases
of computational states,
\begin{equation}
\theta_{\ket{s_1s_2s_3s_4}}
=
\phi_{\ket{s_1s_2s_3s_4}}
-
\sum_{i=1}^{4} s_i\, \phi_{\ket{s_i}},
\end{equation}
where $\phi_{\ket{s_1s_2s_3s_4}}$ is the raw phase of the state component and
$s_i\in\{0,1\}$ denotes the excitation in $\mathrm Q_i$.
Including all relative phases yields the total coherent fidelity $\bar{F}$, and the
total phase-related contribution is $\bar{F}_{P}-\bar{F}$.

To isolate parasitic phase accumulation involving spectator qubits, we evaluate
$\bar{F}_{ZZ}$ by treating the intended controlled phase of the target qubit pair as
ideal and excluding those target-pair controlled phases from the error metric.
For a gate applied to the $\mathrm Q_1$-$\mathrm Q_2$ pair, this corresponds to masking the
four phases $\theta_{\ket{11s_3s_4}}$, which are set by the designed gate operation,
while retaining the phases of the remaining $12$ computational states (including
states such as $\ket{1011}$). These retained phases capture residual $ZZ$ rotations
arising from spectator-spectator and spectator-target pairs.
With this definition, $\varepsilon_{ZZ}=\bar{F}_{P}-\bar{F}_{ZZ}$ quantifies
crosstalk-induced phase errors on non-participating qubits, whereas
$\varepsilon_{\mathrm{gate},ZZ}=\bar{F}_{ZZ}-\bar{F}$ quantifies the residual phase
error associated with the target pair.

The coherent error budget for both gate types is summarized in Fig.~\ref{fig:errorbudget}.
For all qubit pairs, $\varepsilon_{ZZ}$ remains negligible, confirming effective suppression of residual $ZZ$ interactions.
The iSWAP gates are primarily limited by $\varepsilon_{\mathrm{gate},ZZ}$, arising from imperfect cancellation of the controlled phase during the gate; this contribution can be reduced by optimizing the gate duration (see the next subsection).
In contrast, the CZ gates are dominated by leakage errors, with $\varepsilon_{L}$ constituting the largest contribution to the total infidelity.
During the CZ gate, the qubit detuning is close to the anharmonicity to populate the second-excited state of one qubit. This may result in more complex level structures and dynamics in higher-excitation manifolds, increasing leakage. In addition, the idling configuration employed here was optimized to minimize errors in the iSWAP gates and may not be fully optimal for CZ operations.
Further improvements are therefore expected from optimized pulse shaping, frequency allocation, and device-parameter tuning to suppress leakage errors.

\subsection{$ZZ$-free operation of the iSWAP gate}\label{sec:four-qubit-ZZ-free}

\begin{figure}[t]  % figure* → 두 컬럼 전체 폭 사용
    \centering
    \includegraphics[width=\linewidth]{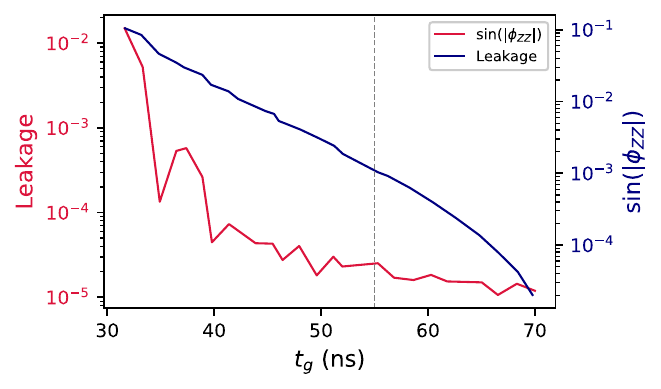}
    \caption{\label{fig:iswapzz} Leakage of the $\ket{1100}$ state (crimson, left axis) and $\sin(|\phi_{ZZ}|)$ (navy, right axis) as functions of the gate duration $t_g$, where $\phi_{ZZ}$ denotes the accumulated phase error of the $\ket{1100}$ state.  The gray dashed line marks the operating point used in Table~\ref{tab:table1} ($t_g=55$\:ns, $\sin(|\phi_{ZZ}|)\approx10^{-3}$). A near–$ZZ$-free condition is obtained around $t_g\approx70$\:ns.
}
\end{figure}

We next examine the operating point required for a $ZZ$-free iSWAP gate in the four-qubit system.
As discussed in Sec.~\ref{sec:gate-operations}, the accumulated controlled phase depends sensitively on the gate duration $t_g$.
By sweeping $t_g$ for an iSWAP gate applied to the $\mathrm{Q}_1$–$\mathrm{Q}_2$ pair, we monitor both the leakage of the $\ket{1100}$ state and its associated phase error $\phi_{ZZ}$ (Fig.~\ref{fig:iswapzz}).
The condition for a $ZZ$-free operation corresponds to $\sin(|\phi_{ZZ}|)=0$.

Although leakage falls below $10^{-4}$ for $t_g \gtrsim 40$\:ns, complete cancellation of the residual phase occurs only near $t_g \approx 70$\:ns.
While this duration minimizes coherent phase errors, the longer gate time increases the error from decoherence.
We therefore choose a compromise duration of $t_g=55$\:ns (gray dashed line), which suppresses the phase error to $\sin(|\phi_{ZZ}|)\approx10^{-3}$ and yields the fidelities reported in Table~\ref{tab:table1}, corresponding to $\varepsilon_{\mathrm{gate},ZZ}\approx4\times10^{-4}$.
To verify that leakage is not the limiting factor, we further evaluate shorter durations of $t_g=45$–$50$\:ns and obtain $\bar{F}_{ZZ}=0.9997\pm6.1\times10^{-5}$, comparable to the optimized values.
This confirms that iSWAP performance is primarily limited by controlled-phase errors rather than leakage.

The longer $ZZ$-free duration compared to the two-qubit case ($\sim45$\:ns) indicates that the zero-crossing point of the residual $ZZ$ coupling $\zeta$ with respect to $\omega_\mathrm{C}$ is shifted.
We attribute this shift to small frequency renormalizations of the activated couplers arising from their mutual hybridization.
Further tuning of coupler parameters or operating frequencies can reduce the required duration for $ZZ$-free operation.

\section{Estimation of incoherent errors from decoherence}
\label{sec:coherence}

\subsection{Two-qubit system}\label{sec:coherence-2q}
To estimate the performance limits set by decoherence, we evaluate the
coherence-limited average gate fidelity using analytic expressions derived
for weak, Markovian noise, as summarized in the Appendix of
Ref.~\cite{2024router}. These formulas capture the leading-order fidelity
reduction arising from independent energy relaxation ($T_1$) and pure
dephasing ($T_\phi$) of the qubits during the gate operation.

For a CZ gate, the dominant dynamics involve transient population transfer
between the computational state $\ket{11}$ and a non-computational state
($\ket{20}$ or $\ket{02}$). As a result, one qubit temporarily occupies the
second excited state with a larger average population; we label this qubit
as $\mathrm{Q}_1$. Under Markovian decoherence, the coherence-limited average
fidelity is
\begin{equation}
\bar F_{\mathrm{CZ}}
=
1
-\frac{1}{2}\frac{t_g}{T_1^{\mathrm{Q}_1}}
-\frac{3}{10}\frac{t_g}{T_1^{\mathrm{Q}_2}}
-\frac{31}{40}\frac{t_g}{T_\phi^{\mathrm{Q}_1}}
-\frac{3}{8}\frac{t_g}{T_\phi^{\mathrm{Q}_2}},
\label{eq:czfidel}
\end{equation}
where $t_g$ is the gate duration and $T_1^{\mathrm{Q}_i}$ and
$T_\phi^{\mathrm{Q}_i}$ denote the relaxation and pure dephasing times of
qubit $\mathrm{Q}_i$. The asymmetric coefficients reflect the unequal population
distribution during the gate.
\begin{figure}[t]
    \centering
    \includegraphics[width=\linewidth]{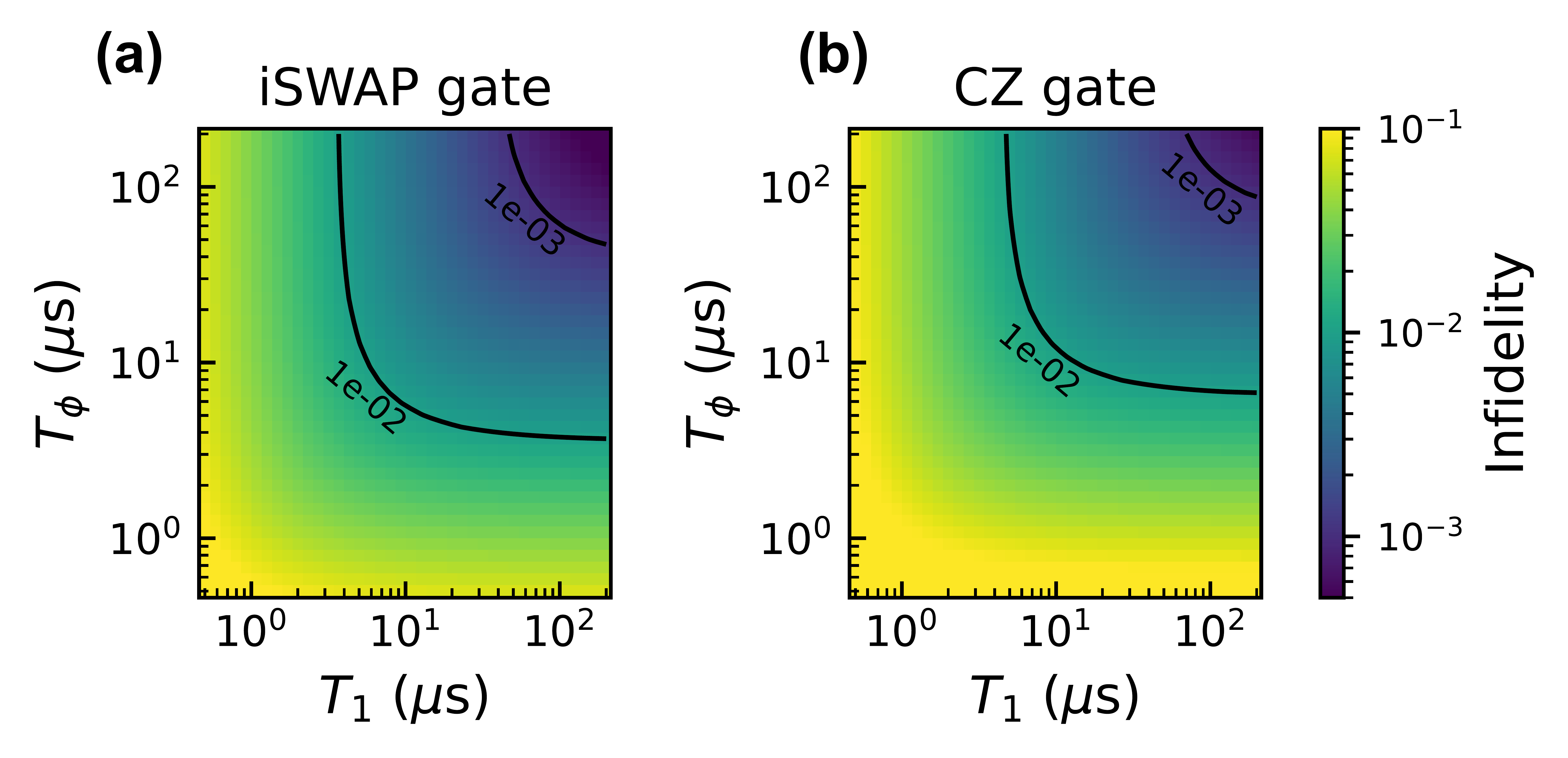}
    \caption{\label{fig:coherence-2q} 
    Estimated coherence-limited total infidelity of (a) the iSWAP gate ($t_g=45$\:ns) and (b) the CZ gate ($t_g=57$\:ns) for an isolated two-qubit register, corresponding to the gate parameters used in Sec.~\ref{sec:gate-operations}.
}
\end{figure}
\begin{figure}[t]
    \centering
    \includegraphics[width=\linewidth]{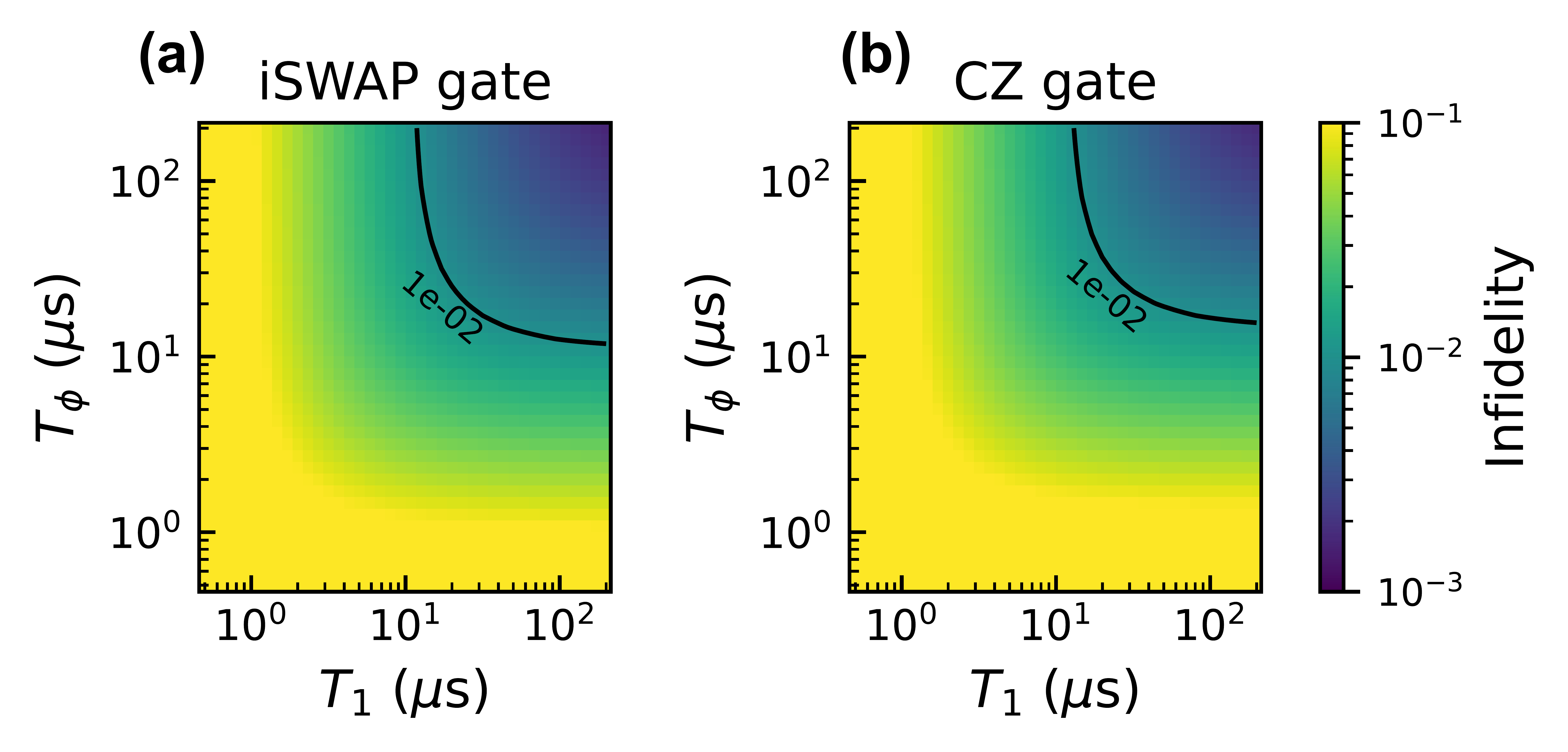}
    \caption{\label{fig:coherence-4q} 
    Estimated coherence-limited total infidelity of (a) the iSWAP gate ($t_g=55.5$\:ns) and (b) the CZ gate ($t_g=60.9$\:ns) applied to the $\mathrm{Q_1}$--$\mathrm{Q_2}$ pair within the full four-qubit register. Gate parameters correspond to Tables~\ref{tab:table1}--\ref{tab:table2}.
}
\end{figure}\label{sec:coherence-4q}
In contrast, the iSWAP gate exchanges population symmetrically between the
$\ket{10}$ and $\ket{01}$ states, leading to identical dynamics for both
qubits. The coherence-limited fidelity therefore takes the symmetric form
\begin{equation}
\bar F_{\mathrm{iSWAP}}
=
1
-\frac{2}{5}t_g
\left(
\frac{1}{T_1^{\mathrm{Q}_1}}
+\frac{1}{T_1^{\mathrm{Q}_2}}
+\frac{1}{T_\phi^{\mathrm{Q}_1}}
+\frac{1}{T_\phi^{\mathrm{Q}_2}}
\right).
\label{eq:iswapfidel}
\end{equation}
These expressions account only for decoherence of the two qubits and neglect
loss in the resonator and couplers, whose populations remain small during
gate operations.

To estimate the total performance, we combine the coherent errors obtained
from numerical simulation with the incoherent errors predicted above.
Using the values obtained in Sec.~\ref{sec:gate-operations} [$(t_g,\bar{F})$ of $(45\:\mathrm{ns}, 99.995\%)$ for the iSWAP gate and $(57\:\mathrm{ns},99.998\%)$ for the CZ gate], and assuming identical coherence times
for both qubits, we evaluate the total infidelity as a function of $T_1$
and $T_\phi$ (Fig.~\ref{fig:coherence-2q}). Gate fidelities above $99.9\:\%$ can be achieved with $T_1, T_\phi\approx 100\:\mu$s.

\subsection{Four-qubit system}\label{sec:coherence-4q}

We also estimate the coherence-limited fidelity of selective gates discussed in Sec.~\ref{sec:four-qubit}. The gate operations can be expressed as the tensor product of two identity matrix for each spectator qubit and the unitary of a two-qubit gate, either $\mathrm{iSWAP}$ or $\mathrm{CZ}$ acting on the target pair, $\mathrm{CZ} \otimes I \otimes I$ for instance. In such cases, decoherence of the spectator qubits during a selective gate operation that is ideally intended to apply the identity operation degrades the overall performance evaluated on the complete computational basis. Hence, decoherence of spectator qubits must be taken into account to quantify the coherence-limited fidelity.

From Ref.~\cite{generalize-limit}, fidelity of an operation limited by the uncorrelated energy relaxation and pure dephasing on all $n$ qubits in the system is given by
\begin{equation}
\bar{F} = 1-\frac{d}{2(d+1)}t_g\sum^n_{k=1}\left(\frac{1}{T_1^{\mathrm{Q}_k}} +\frac{1}{T_\phi^{\mathrm{Q}_k}}\right),\label{eq:generalized-f}
\end{equation}
with $d=2^n$ and the operation time $t_g$. The condition for this equation to hold is that the intended dynamics of the operation should take place within the computational subspace, therefore including iSWAP gates and identity operations. For $n=2$, Eq.~\eqref{eq:generalized-f} reproduces Eq.~\eqref{eq:iswapfidel}. The iSWAP operation with a duration $t_g$ in a four-qubit system can be written as 
\begin{equation}
\bar{F} = 1-\frac{8}{17}t_g\sum^n_{k=4}\left(\frac{1}{T_1^{\mathrm{Q_k}}} +\frac{1}{T_\phi^{\mathrm{Q_k}}}\right).
\end{equation}

For CZ gates that leave the computational subspace during its operation, the coefficients on $\tau/{T_1}$ and $\tau/{T_\phi}$ are asymmetric. These coefficients can be inferred from the fidelity estimation of simultaneous CZ gates discussed in Ref.~\cite{simul-zz}. Combining with the contributions of identity operations on the spectator qubits, the coherence-limited fidelity of a CZ operation in a four-qubit system is then given by
\begin{align}
\bar{F} &= 1-\frac{10}{17}\frac{t_g}{T_1^{\mathrm{Q}_a}}-\frac{6}{17}\frac{t_g}{T_1^{\mathrm{Q}_b}}-\frac{245}{272}\frac{t_g}{T_\phi^{\mathrm{Q}_a}}\nonumber\\
&\qquad -\frac{117}{272}\frac{t_g}{T_\phi^{\mathrm{Q}_b}}
-\frac{8}{17}\sum_{k\ne a,b}\left(\frac{t_g}{T_1^{\mathrm{Q}_k}} +\frac{t_g}{T_\phi^{\mathrm{Q}_k}}\right),
\end{align}
where the CZ gate acting on the $\mathrm{Q}_a$-$\mathrm{Q}_b$ pair, and $\mathrm{Q}_a$ is the qubit temporarily occupying its second-excited state during the gate. 

Using the representative gate durations in Tables~\ref{tab:table1}--\ref{tab:table2} [$(t_g,\bar{F})$ of $(55.5\:\mathrm{ns}, 99.94\%)$ for the iSWAP gate and $(60.9\:\mathrm{ns},99.95\%)$ for the CZ gate between Q$_1$ and Q$_2$], and assuming identical coherence times for both qubits, we evaluate the total infidelity as a function of $T_1$ and $T_\phi$ (Fig.~\ref{fig:coherence-4q}). For $T_1, T_\phi \approx 30\:\mu$s, fidelities of $99\%$ are expected, and $T_1, T_\phi \approx 150\:\mu$s result in fidelities approaching $99.8\%$. Above $300\:\mu$s relaxation and pure dephasing time is required to achieve $99.9\%$.

For reference, we mimic an experiment that measures the two-qubit gate fidelity for each qubit pair, while rest of the qubits are initialized in their ground states in a multi-qubit device. We confine the computational subspace to only two qubits activated for applying CZ gates. The operation of a gate is $\mathrm{CZ}$ for two qubits, instead of the four-qubit process $\mathrm{CZ}\otimes I \otimes I$. For example, only the four $\ket{s_1s_200}$ states are considered to calculate the fidelity of the CZ gate that acts on the $\mathrm{Q_1}$-$\mathrm{Q_2}$ pair. We repeat such fidelity evaluations for all six qubit pairs and obtain an average fidelity of $0.99978\pm 9.31\times10^{-5}$, higher than the case of considering the complete computational subspace of the four-qubit register. This implies that measurements over the complete computational manifold are required to properly capture leakage and qubit crosstalk. Since the computational subspace is truncated to two qubits, we use Eqs.~\eqref{eq:czfidel}--\eqref{eq:iswapfidel} for incoherent errors. For such a setup, $T_1, T_\phi \approx 150\:\mu$s can achieve $99.9\%$ fidelity.


\nocite{}

\bibliography{apssamp}% Produces the bibliography via BibTeX.

\end{document}